\documentclass[final,3p,times,nopreprintline]{elsarticle}

\usepackage{amsmath}
\usepackage{amssymb}
\usepackage{graphicx}
\usepackage{booktabs}

\biboptions{sort&compress}

\usepackage{geometry}
\AtBeginDocument{\fontsize{11pt}{16pt}\selectfont}

\usepackage{titlesec}

\titleformat{\section}
  {\normalfont\bfseries\fontsize{12pt}{16pt}\selectfont}
  {\thesection}
  {0.8em}
  {}

\titleformat{\subsection}
  {\normalfont\bfseries\fontsize{11pt}{16pt}\selectfont}
  {\thesubsection}
  {0.8em}
  {}

\titleformat{\subsubsection}
  {\normalfont\bfseries\fontsize{11pt}{16pt}\selectfont}
  {\thesubsubsection}
  {0.8em}
  {}

\usepackage[section]{placeins}

\usepackage[
    colorlinks=true,
    linkcolor=blue,
    urlcolor=blue,
    citecolor=blue
]{hyperref}

\begin{document}

\begin{frontmatter}

\title{NEP-kappa 2.0: A unified framework for phonon transport and nonequilibrium thermalization with machine-learning interatomic potentials}

\author[1]{Shixian Liu\fnref{equal}}
\author[2]{Bin Liu\fnref{equal}}
\author[1]{Fei Yin}
\author[3]{Gang Wang}
\author[1]{Yiming Dong}
\author[4]{Yangjun Qin}
\author[1]{Alexandra V. Kosyakova}
\author[1]{Qiang Xian}
\author[1]{Alexander A. Barinov\corref{cor1}}
\ead{barinov@bmstu.ru}
\author[3]{Ke Xu\corref{cor1}}
\ead{kickhsu@gmail.com}
\author[5]{Gang Zhang\corref{cor1}}
\ead{gangzhang@bit.edu.cn}

\cortext[cor1]{Corresponding authors}
\fntext[equal]{These authors contributed equally to this work.}

\address[1]{
Department of Thermophysics,
Bauman Moscow State Technical University,
Moscow 105005, Russia
}

\address[2]{
Key Laboratory for Thermal Science and Power Engineering of Ministry of Education,
Department of Engineering Mechanics,
Tsinghua University,
Beijing 100084, China
}

\address[3]{
College of Physical Science and Technology,
Bohai University,
Jinzhou, China
}

\address[4]{
State Key Laboratory of Fluid Power and Mechatronic System,
School of Mechanical Engineering,
Zhejiang University,
Hangzhou 310058, China
}

\address[5]{
School of Materials Science and Engineering,
Beijing Institute of Technology,
Beijing 100081, China
}

\begin{abstract}
We present NEP-kappa 2.0, a unified framework that integrates density-functional theory, material-specific machine-learning interatomic potentials, and pretrained force models for advanced phonon calculations. A common workflow enables harmonic lattice dynamics, quasi-harmonic and stochastic self-consistent phonon treatments, three- and four-phonon transport, Wigner population--coherence transport, and time-dependent phonon dynamics. Applications to 3C-SiC, BAs, and Cs$_2$AgInCl$_6$ demonstrate consistent force-model comparisons, finite-temperature renormalization, higher-order scattering, and intermode coherence. Beyond steady-state transport, mode-resolved simulations of selectively excited 3C-SiC reveal that half of the injected optical-phonon energy reaches acoustic modes within approximately 3.75~ps, primarily through one-optical--two-acoustic scattering. Acoustic branches exhibit response times differing by a factor of five, with competing scattering channels producing a transient energy overshoot. Channel-resolved entropy production and persistent nonthermal occupations distinguish interbranch energy redistribution from intrabranch thermalization. The framework connects equilibrium phonon transport with microscopic nonequilibrium energy dynamics.
\end{abstract}


\end{frontmatter}

\clearpage


\section{Introduction}

First-principles lattice dynamics has established a quantitative route from interatomic interactions to phonon thermal transport in crystalline materials. Second-order force constants (FC2) determine phonon frequencies, eigenvectors, and group velocities~\cite{togo2015_phonon,togo2023_phonopy}, while third-order force constants (FC3) provide the cubic interactions required for three-phonon scattering and solutions of the phonon Boltzmann transport equation (BTE)~\cite{omini1995,broido2007,esfarjani2011,chaput2013,li2014_shengbte,togo2015_lifetime}. Thermal conductivity then follows from the combined effects of modal heat capacities, phonon velocities, scattering phase space, and anharmonic interaction strengths~\cite{lindsay2008}. Extending the expansion to fourth-order force constants (FC4) introduces four-phonon processes, which can substantially modify phonon lifetimes and thermal conductivity when three-phonon channels are restricted or at elevated temperatures~\cite{feng2016,feng2018,yang2019_fourphonon,han2022_fourphonon,li_higher-order-phonon_2026}. The cost of generating the required force information, however, increases rapidly with interaction order and configurational complexity.

The same difficulty becomes more pronounced when the phonon spectrum itself evolves with temperature or when transport moves beyond the conventional steady-state population picture. Thermal expansion can be treated within the quasi-harmonic approximation (QHA)~\cite{togo2010_qha}, whereas intrinsic anharmonic renormalization requires sampling the finite-temperature potential-energy surface, as in temperature-dependent effective-potential methods~\cite{hellman2013_tdep,hellman2013_fc3}, self-consistent phonon theory~\cite{tadano2015,tadano2018}, and the stochastic self-consistent harmonic approximation (SSCHA)~\cite{errea2014,monacelli2021}. In crystals with dense vibrational spectra, the Wigner formulation further accounts for transport associated with coherences between different phonon branches~\cite{simoncelli2019,simoncelli2022}. Nonequilibrium problems require the occupations themselves to be followed in time, allowing phonon--phonon collisions to redistribute energy after selective excitation or external energy deposition~\cite{maldonado2020,caruso2022}. These extensions access increasingly rich phonon physics, but they also require large numbers of force evaluations or increasingly demanding higher-order derivatives of the potential-energy surface.

Machine-learning interatomic potentials (MLIPs) provide an efficient alternative to repeated first-principles force calculations. Representative approaches include systematically improvable moment tensor potentials (MTPs)~\cite{shapeev2016_mtp}, Gaussian approximation potentials (GAPs)~\cite{bartok2010}, neuroevolution potentials (NEPs)~\cite{fan2021}, and equivariant models such as MACE~\cite{batatia2022_mace}. Their applications in phonon thermal transport range from harmonic and anharmonic lattice dynamics~\cite{babaei2019,ladygin2020,yin2026_nepkappa} to equilibrium molecular dynamics and Green--Kubo calculations~\cite{qian2019,xu2025gpumd,korotaev2019,qin2026_mlp_review}, as well as force-constant extraction followed by BTE calculations~\cite{mortazavi2021,choi2022_nn,togo2024_otf}. MLIPs have also been applied to thermal transport in complex compounds~\cite{korotaev2019}, interfaces~\cite{wyant2021_interface}, and confined nanostructures~\cite{liu2025_nep_specific_heat}. Recent work on cove-edged graphene nanoribbons has further demonstrated how edge-induced vibrational confinement can weaken collective phonon transport~\cite{yin_cove_2027}. Further developments include higher-order phonon anharmonicity and four-phonon transport~\cite{liu2021_bas_fourphonon,ouyang2022,tang2023_hbn,zhao2026_in2se3,li_higher-order-phonon_2026}, stochastic finite-temperature calculations~\cite{belli2025_sscha_mlip}, and high-throughput phonon calculations using transferable force models~\cite{lee2025_highthroughput,srivastava2024}. MLIPs therefore offer more than an acceleration of molecular dynamics: they can act as efficient providers of the forces and derivatives required throughout a lattice-dynamical workflow.

Using an MLIP as a force engine for lattice dynamics also differs in an important way from using it directly in molecular dynamics. Residual force errors in MLIP-based MD have been shown to behave as additional perturbations to the atomic forces, increasing low-frequency phonon scattering and producing a systematic underestimation of lattice thermal conductivity in several representative materials~\cite{wu2024_forceerror}. In a force-constant-based BTE treatment, the learned potential is instead sampled to construct the derivatives of the potential-energy surface, so that force errors do not accumulate along a dynamical trajectory in the same manner. 
Errors in the MLIP remain relevant through the resulting FC2, FC3, and FC4 and may lead to either over- or underestimation of transport properties, but they do not introduce the same trajectory-level artificial scattering mechanism as in MLIP-based MD. This distinction provides an additional motivation for combining the efficiency of MLIPs with mode-resolved lattice dynamics and quantum phonon statistics.

The neuroevolution potential (NEP) is particularly attractive for force-intensive calculations because of its efficient GPU implementation~\cite{fan2021}. Its implementation within the GPUMD package has recently been further developed into GPUMD 4.0, which provides a high-performance platform for a broad range of atomistic simulations with machine-learned potentials~\cite{xu2025gpumd}. The recent NEP89 model further extends the same architecture toward transferable interatomic modeling across a broad chemical space~\cite{liang2026_nep89}. In the original NEP-kappa framework, NEP forces were coupled with compressed-sensing force-constant extraction and phonon-transport calculations, including the Wigner formulation~\cite{yin2026_nepkappa}. Transferable equivariant models such as MACE provide a complementary route in which a pretrained model can be applied without constructing a dedicated potential for each material~\cite{batatia2022_mace,batatia2025_foundation,kaplan2025_matpes}. Their vibrational accuracy nevertheless remains property dependent: good agreement in energies and forces does not necessarily guarantee equally accurate phonon spectra~\cite{loew2025}, which has motivated phonon-oriented fine tuning using Hessians or force constants~\cite{koker2026_pft}. Recent thermal-transport packages have likewise begun to incorporate both pretrained and material-specific MLIPs into lattice-dynamical calculations~\cite{kaldo2026}.

A separate issue concerns the consistency of the computational workflow itself. Modern phonon calculations commonly combine specialized tools for structural preparation, force-constant construction, finite-temperature sampling, and transport solutions. Such modularity is useful, but direct comparisons can become ambiguous when different force models are accompanied by changes in structural references, atom mapping, force-constant conventions, symmetry treatment, reciprocal-space meshes, broadening parameters, or transport solvers. A more fundamental inconsistency arises if FC2, FC3, and FC4 are generated from different interatomic descriptions or structural states, because the harmonic spectrum and anharmonic interaction matrix elements then no longer derive from the same potential-energy surface. Differences in a final thermal conductivity may consequently contain contributions from both the force model and the surrounding numerical workflow. A common computational layer that preserves structural definitions, force-constant conventions, reciprocal-space sampling, and observable definitions therefore makes model comparisons more transparent and reduces workflow-dependent sources of uncertainty.

The need for such a framework becomes especially clear for nonequilibrium phonon dynamics, where conventional lifetimes or steady-state scattering rates do not specify how energy actually flows between phonon subsystems. Ultrafast external excitation has also emerged as a route to actively modify thermal transport, with recent calculations predicting laser-induced switching of lattice thermal conductivity on femtosecond time scales~\cite{zhang2026_mnse}. Anharmonic decay of optical phonons into lower-energy modes in cubic SiC has long been inferred from Raman linewidth measurements~\cite{olego1982_sic_decay}, and nonequilibrium longitudinal-optical phonons have been observed under high-field conditions in SiC devices~\cite{matulionis2004_sic_hotphonon}. Mode-selective phonon excitation has also been demonstrated experimentally in SiC using mid-infrared irradiation~\cite{yoshida2013_sic_excitation}. These experimental and theoretical studies establish the relevance of optical-phonon relaxation to hot-phonon dissipation and selectively driven lattice dynamics. They do not, however, directly resolve the transient, mode-resolved energy flow among individual phonon branches or distinguish the signed contributions of different phonon--phonon scattering channels. A time-dependent BTE can address this gap by following the phonon occupations explicitly in time and resolving the microscopic pathways of energy redistribution.

Here we introduce NEP-kappa 2.0, a unified framework in which first-principles calculations, material-specific MLIPs, and pretrained interatomic models serve as interchangeable force engines for advanced phonon calculations. A common computational interface connects these force descriptions to harmonic lattice dynamics, QHA and SSCHA finite-temperature phonons, three- and four-phonon transport, Wigner population--coherence transport, and time-dependent phonon population dynamics. The framework is examined across several representative materials. In 3C-SiC, DFT, NEP, and MACE are compared within the same harmonic and three-phonon transport workflow, followed by finite-temperature QHA and SSCHA calculations. BAs is used to validate the higher-order branch with combined three- and four-phonon transport, while Cs$_2$AgInCl$_6$ illustrates Wigner transport in a dense vibrational spectrum. The time-dependent formulation is then applied to selectively excited 3C-SiC to resolve the transfer of excess optical-phonon energy into the acoustic subsystem and to identify the scattering channels responsible for the net energy flow. The population-transport branch also retains the full linearized collision operator and resolves Normal and Umklapp scattering separately, providing the microscopic information required for analyses of phonon hydrodynamics. In this way, NEP-kappa 2.0 combines a consistent MLIP-enabled workflow with access to phonon observables ranging from equilibrium transport coefficients to transient, mode-resolved energy redistribution.

\section{Results}

\subsection{Overview of the NEP-kappa 2.0 framework}

NEP-kappa 2.0 separates force evaluation from lattice-dynamical and transport calculations through a modular architecture. Energies and atomic forces can be supplied by density-functional theory (DFT), a material-specific machine-learning interatomic potential such as NEP~\cite{fan2021}, or a pretrained model such as MACE~\cite{batatia2022_mace,kaplan2025_matpes}. These force engines support structural relaxation, force-constant generation, and finite-temperature sampling within a common workflow. Building on the original NEP-based implementation~\cite{yin2026_nepkappa}, NEP-kappa 2.0 extends the framework to interchangeable force engines while retaining the same downstream physical formulations. The calculations presented here demonstrate this capability using both NEP and MACE.

\begin{figure}[htbp]
    \centering
    \includegraphics[width=0.7\linewidth]{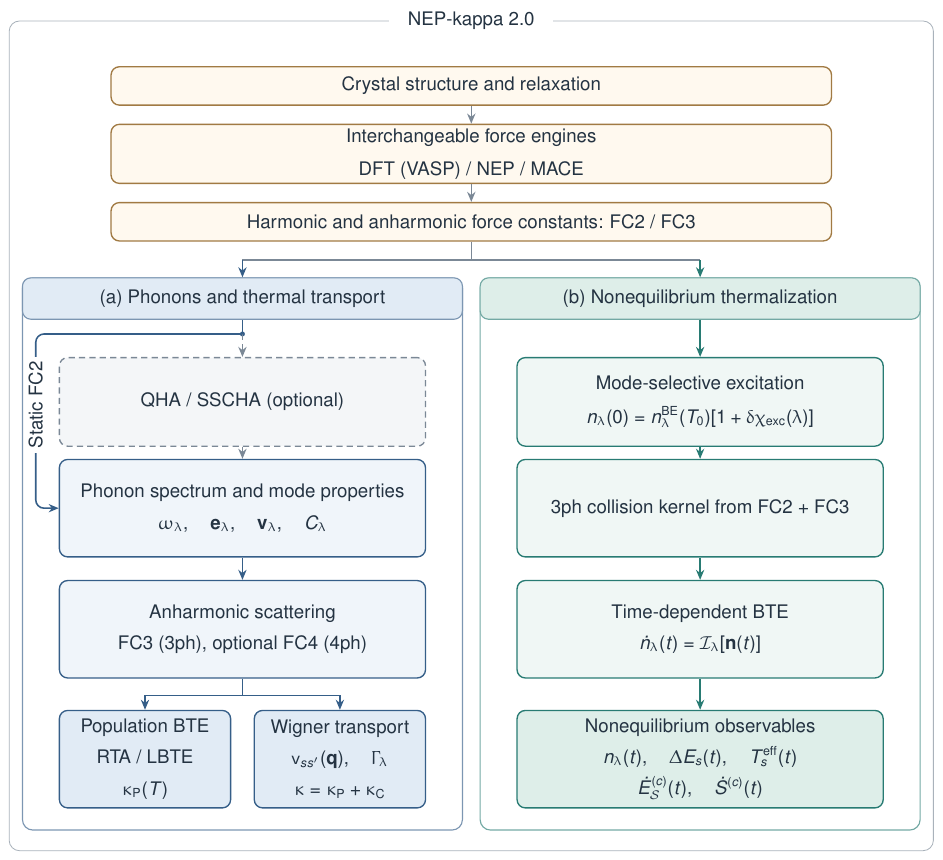}
    \caption{Overview of the NEP-kappa 2.0 framework. Shared structural preparation and force evaluation feed two calculation routes. (a) Phonon properties and thermal transport. Static FC2 or optional QHA/SSCHA treatments determine the phonon spectrum and mode properties. FC3 and optional FC4 provide three- and four-phonon scattering, respectively. Population transport is evaluated using RTA or iterative LBTE; Wigner transport resolves the population and coherence contributions, $\kappa=\kappa_{\mathrm P}+\kappa_{\mathrm C}$. (b) Nonequilibrium thermalization. Mode-selective excitation initializes a time-dependent BTE calculation, yielding mode occupations, branch excess energies, effective branch temperatures, and channel-resolved energy transfer and entropy production. The dashed box and arrows denote optional temperature treatments, whereas the solid bypass uses static FC2. Here, $\lambda=(\mathbf q,s)$ denotes a phonon mode, $\mathcal S$ a selected set of modes, and $c$ a class of scattering processes.}
    \label{fig:framework}
\end{figure}

Figure~\ref{fig:framework} summarizes the shared preparation steps and the two principal calculation routes. The first route addresses phonon properties and thermal transport. Phonopy and phono3py provide harmonic lattice dynamics and three-phonon transport calculations~\cite{togo2023_phonopy}, while FourPhonon provides four-phonon calculations based on fourth-order force constants (FC4)~\cite{han2022_fourphonon}. Temperature effects can be included through QHA, which determines the equilibrium volume as a function of temperature, and SSCHA, which provides auxiliary effective second-order force constants at a specified temperature and volume~\cite{togo2010_qha,monacelli2021}. The two treatments can also be combined sequentially by performing SSCHA at the QHA equilibrium volume. Thermal conductivity is evaluated using the population BTE or the Wigner formulation, with the latter accounting for both population and coherence contributions~\cite{simoncelli2019,simoncelli2022}. The second route addresses nonequilibrium thermalization: a time-dependent BTE evolves mode occupations from a prescribed initial excitation using a three-phonon collision kernel. This route does not require a preceding thermal-conductivity calculation.

A common interface handles input preparation and consistency checks across these routes. Changing the force engine does not change the downstream transport formulation, allowing different force models to be compared within the same theoretical treatment. For the comparisons reported here, atom ordering, crystal-cell definitions, units, force-constant conventions, reciprocal-space sampling, and observable definitions are matched across calculations. Numerical convergence and physical validation are assessed separately for each calculation.

The supercell used to construct force constants and the reciprocal-space $q$ mesh used for transport integration are converged independently because they control different sources of numerical error. The supercell determines the spatial range of interactions that can be resolved in the force-constant representation, whereas the $q$ mesh controls the resolution of Brillouin-zone integration. Separating these convergence tests helps distinguish differences between force models from finite-supercell and sampling effects when comparing DFT and MLIP results.

The following sections demonstrate the framework through representative applications covering harmonic and anharmonic phonon transport, finite-temperature renormalization, higher-order phonon scattering, Wigner population--coherence transport, and nonequilibrium phonon thermalization.

\subsection{MLIP-enabled phonon transport workflow in 3C-SiC}

\begin{figure}[h]
    \centering
    \includegraphics[width=\linewidth]{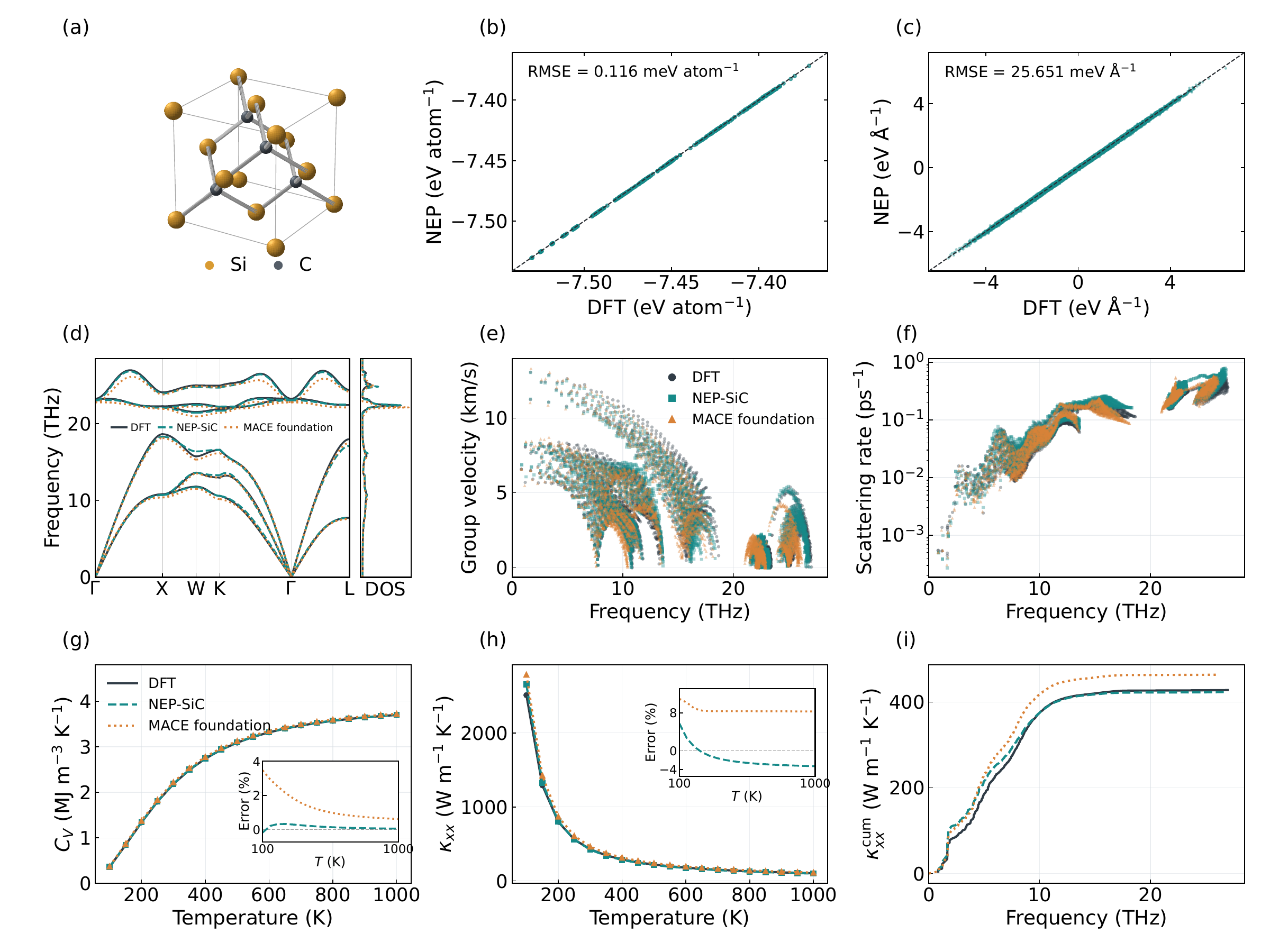}
    \caption{
    Phonon transport in 3C-SiC calculated from DFT, a material-specific NEP-SiC potential, and the pretrained MACE-MATPES-PBE-0 model.
    (a) Crystal structure.
    (b,c) NEP-SiC energies and Cartesian force components compared with the DFT training data; dashed lines indicate exact agreement.
    (d) Phonon dispersion along $\Gamma$--X--W--K--$\Gamma$--L and the corresponding density of states.
    (e) Mode-resolved group-velocity magnitudes.
    (f) Three-phonon scattering rates at 300~K.
    (g) Volumetric heat capacity.
    (h) RTA lattice thermal conductivity.
    (i) Frequency-resolved cumulative thermal conductivity at 300~K.
    Insets in (g,h) show the signed relative deviation from DFT,
    $100(X_{\mathrm{model}}-X_{\mathrm{DFT}})/X_{\mathrm{DFT}}$.
    The marker legend in (e) also applies to (f), and the line legend in (g) applies to (g--i).
    }
    \label{fig:sic}
\end{figure}

3C-SiC is used as a benchmark for the complete force-to-transport workflow and for comparing different force engines under identical downstream numerical settings. Figure~\ref{fig:sic} presents results obtained from DFT, a material-specific NEP-SiC potential, and the pretrained MACE-MATPES-PBE-0 model. For NEP-SiC, the energy and force predictions reproduce the DFT training data with root-mean-square errors of 0.116~meV\,atom$^{-1}$ and 25.651~meV\,\AA$^{-1}$, respectively [Fig.~\ref{fig:sic}(b,c)]. The three force descriptions are then used without changing the subsequent lattice-dynamical and transport workflow.

At the harmonic level, both MLIPs reproduce the main features of the DFT phonon spectrum [Fig.~\ref{fig:sic}(d)]. The agreement is especially close for NEP-SiC, whereas MACE shows somewhat larger shifts in parts of the optical branches. These differences have only a limited effect on the volumetric heat capacity [Fig.~\ref{fig:sic}(g)], which depends primarily on the phonon frequencies and Bose--Einstein occupations and therefore remains close to the DFT result over the temperature range considered.

Differences become more pronounced in quantities that depend on the dispersion derivatives and anharmonic interactions. The group velocities in Fig.~\ref{fig:sic}(e) retain the overall spectral structure obtained from DFT but show a broader model dependence than the frequencies themselves. A larger spread is evident in the three-phonon scattering rates [Fig.~\ref{fig:sic}(f)], which depend on the harmonic frequencies and eigenvectors together with the cubic interaction matrix elements. The larger differences in scattering rates highlight their sensitivity to anharmonic interactions beyond the harmonic phonon spectrum.

The resulting differences are reflected in the thermal conductivity. At 300~K, the RTA conductivities are approximately 428, 423, and 464~W\,m$^{-1}$\,K$^{-1}$ for DFT, NEP-SiC, and MACE, respectively. Relative to DFT, the deviations are about $-1.2\%$ for NEP-SiC and $+8.4\%$ for MACE. The temperature dependence in Fig.~\ref{fig:sic}(h) further shows that these deviations are not described by a single constant offset.

The cumulative conductivity in Fig.~\ref{fig:sic}(i) clarifies the spectral origin of the remaining differences. NEP-SiC closely follows the DFT accumulation over most of the heat-carrying frequency range, whereas MACE develops a larger contribution at intermediate frequencies and consequently reaches a higher total conductivity. 

\subsection{Finite-temperature phonon renormalization and transport in 3C-SiC}

\begin{figure}[h]
    \centering
    \includegraphics[width=\linewidth]{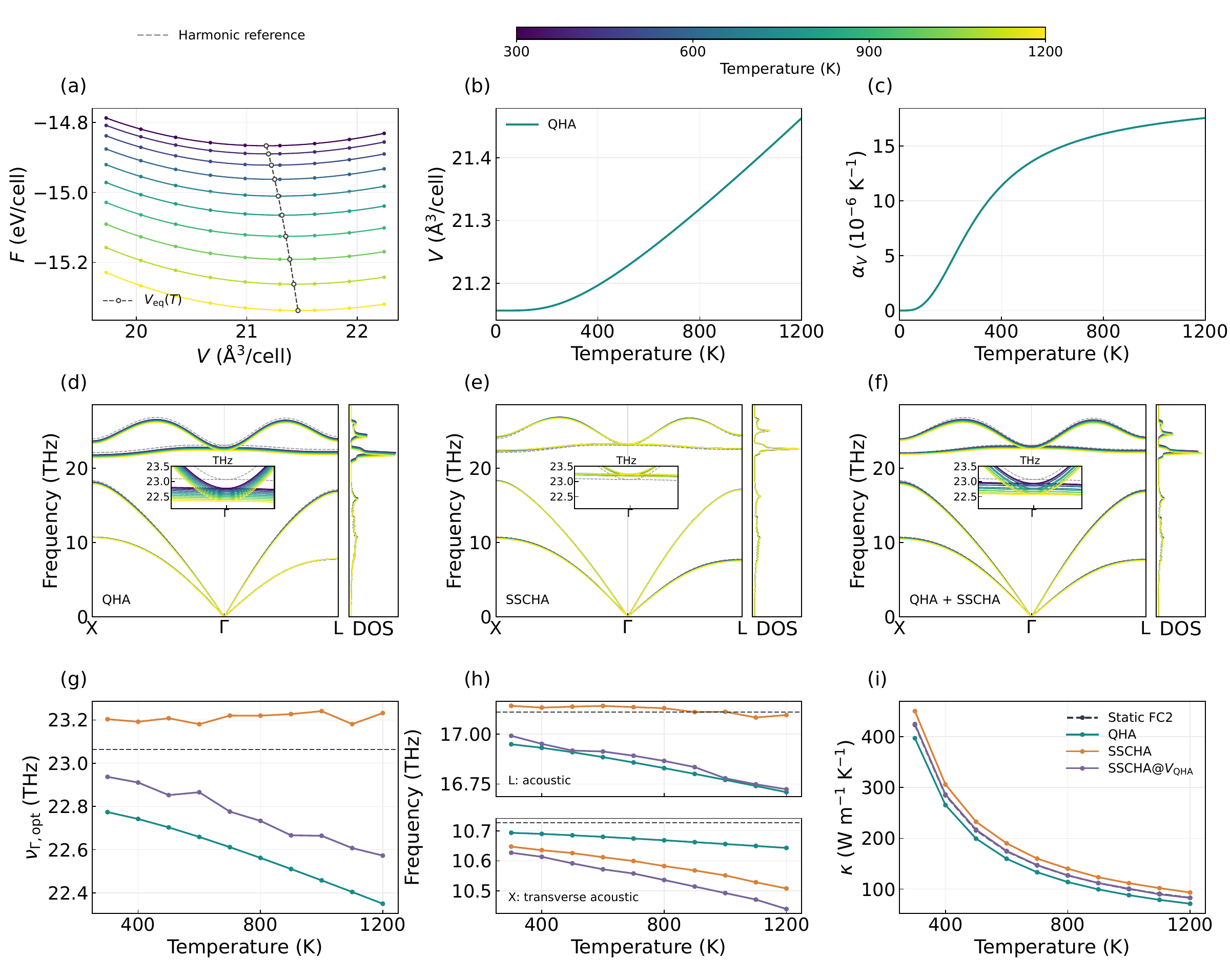}
    \caption{
    Finite-temperature phonon properties and thermal transport of 3C-SiC calculated with the NEP-SiC potential.
    (a) Helmholtz free energy as a function of primitive-cell volume from 300 to 1200~K. Solid curves are Vinet equation-of-state fits, and open circles indicate the equilibrium volumes.
    (b) QHA equilibrium volume and
    (c) volumetric thermal-expansion coefficient.
    (d--f) Temperature-dependent phonon dispersions and densities of states obtained from QHA, fixed-volume SSCHA, and sequential QHA+SSCHA, respectively. Gray dashed lines indicate the static harmonic reference, and the common temperature scale applies to (a) and (d--f).
    (g) Mean frequency of the optical triplet at $\Gamma$.
    (h) Representative acoustic frequencies at the L (top) and X (bottom) points.
    (i) RTA lattice thermal conductivity calculated using static FC2, QHA-renormalized FC2, fixed-volume SSCHA FC2, and SSCHA FC2 evaluated at $V_{\mathrm{QHA}}(T)$.
    }
    \label{fig:sic_temperature}
\end{figure}

The finite-temperature capabilities of NEP-kappa 2.0 are examined using the same NEP-SiC potential. Here, the force model is used not only for displaced structures but also for volume-dependent and stochastic finite-temperature calculations. Three treatments are compared in Fig.~\ref{fig:sic_temperature}: QHA, which accounts for thermal expansion; fixed-volume SSCHA, which isolates intrinsic anharmonic renormalization; and a sequential QHA+SSCHA procedure in which SSCHA is performed at the temperature-dependent QHA equilibrium volume.

The QHA calculations show the expected thermal expansion of 3C-SiC. The free-energy minima shift toward larger volumes with increasing temperature [Fig.~\ref{fig:sic_temperature}(a)], giving the monotonic volume increase in panel (b). The volumetric expansion coefficient rises rapidly at lower temperatures and becomes more weakly temperature dependent at higher temperature [panel (c)]. This expansion is accompanied by an overall softening of the phonon spectrum, most clearly seen in the optical branches of Fig.~\ref{fig:sic_temperature}(d).

Intrinsic anharmonicity produces a different, strongly mode-dependent response. At fixed volume, the SSCHA spectrum remains close to the static harmonic result over much of the Brillouin zone, while the optical triplet near $\Gamma$ shifts to higher frequency [Fig.~\ref{fig:sic_temperature}(e)]. The contrast with QHA is evident in Fig.~\ref{fig:sic_temperature}(g): thermal expansion softens these optical modes, whereas intrinsic anharmonicity produces a weak hardening. The two effects therefore compete for the $\Gamma$-point optical branch.

This competition is not universal across the spectrum. For the selected acoustic modes at L, the fixed-volume SSCHA correction is comparatively small and partly compensates the QHA softening. At X, however, the intrinsic anharmonic correction also lowers the transverse-acoustic frequencies, so that both effects act in the same direction [Fig.~\ref{fig:sic_temperature}(h)]. The renormalization is strongly mode dependent, with competing shifts near $\Gamma$ and reinforcing softening for selected acoustic modes at X.

The sequential QHA+SSCHA calculation evaluates the intrinsic anharmonic renormalization at the thermally expanded structure rather than combining frequency shifts obtained at different volumes. As shown in Fig.~\ref{fig:sic_temperature}(f), this produces a spectrum that reflects both effects within the same structural reference. For the $\Gamma$-point optical triplet, the resulting frequencies lie between the QHA and fixed-volume SSCHA limits [Fig.~\ref{fig:sic_temperature}(g)], consistent with their opposing contributions. The zone-boundary acoustic modes retain the more complex mode-dependent behavior seen in panel (h).

These differences carry over to thermal transport. QHA lowers the RTA conductivity relative to the static-FC2 calculation, whereas fixed-volume SSCHA increases it over the temperature range considered [Fig.~\ref{fig:sic_temperature}(i)]. The sequential QHA+SSCHA result lies between the two. This behavior cannot be inferred from the shift of any single branch, because replacing the static FC2 with the finite-temperature effective FC2 changes the phonon frequencies and eigenvectors together with the group velocities, scattering phase space, and projection of the anharmonic interaction matrix elements.

\subsection{Higher-order anharmonic transport in BAs}

BAs is used to test the higher-order anharmonic branch of NEP-kappa 2.0, extending the workflow from FC3 and three-phonon scattering to FC4 and four-phonon transport. Previous studies have investigated fully iterative solutions of the phonon BTE including both three- and four-phonon scattering in BAs \cite{ravichandran2019nonmonotonic,dai2025machine}. Here, we extend this approach to a unified comparison of DFT, material-specific NEP, and pretrained MACE force models under identical transport settings.

\begin{figure}[h]
    \centering
    \includegraphics[width=\linewidth]{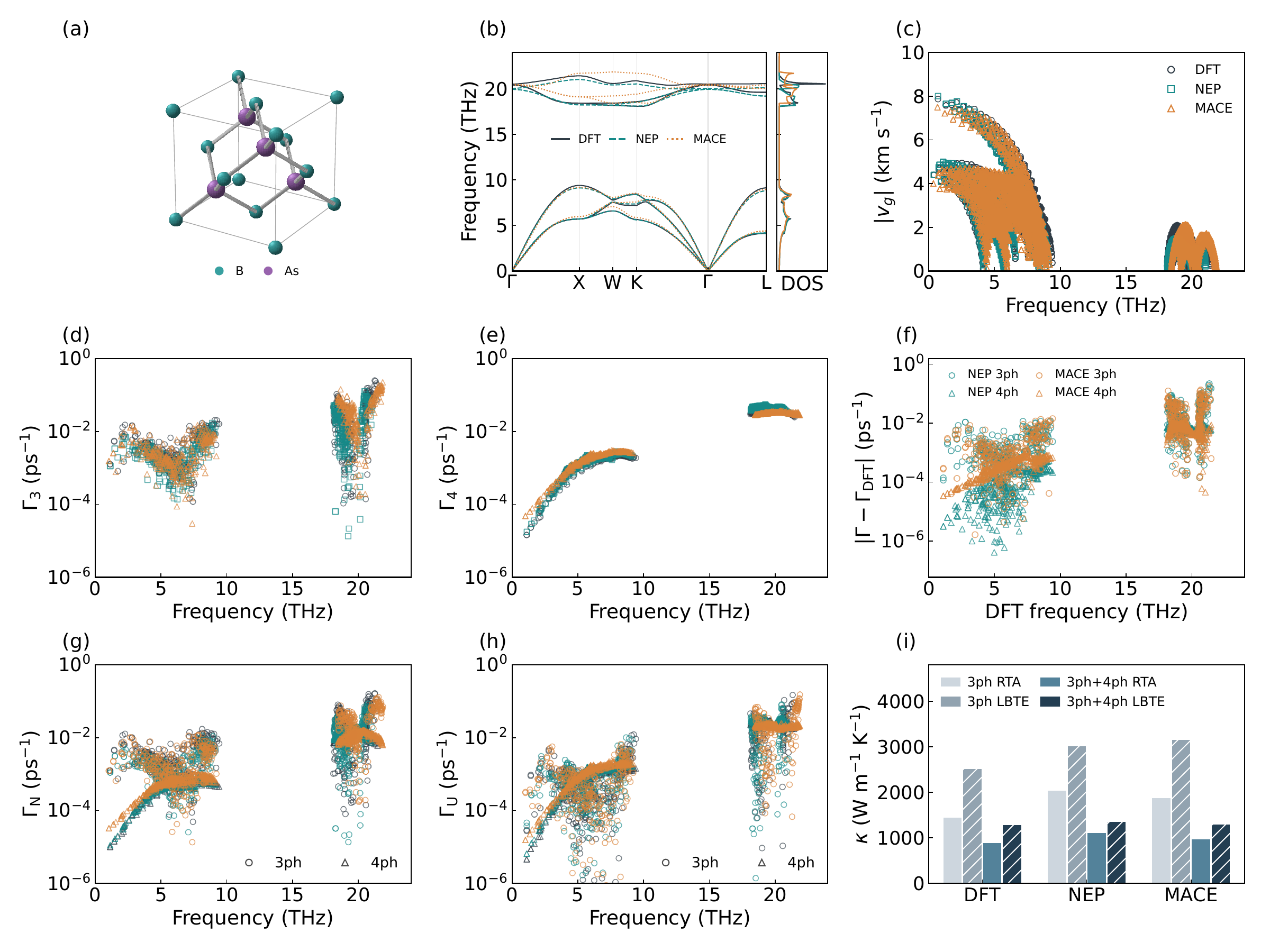}
    \caption{
    Three- and four-phonon transport in BAs calculated using DFT, NEP, and MACE.
    (a) Crystal structure.
    (b) Phonon dispersion and density of states.
    (c) Mode-resolved group velocities.
    (d,e) Three- and four-phonon scattering rates at 300~K.
    (f) Absolute deviations of the NEP and MACE scattering rates from DFT,
    $|\Gamma-\Gamma_{\mathrm{DFT}}|$.
    (g,h) Normal and Umklapp contributions to the three- and four-phonon scattering rates.
    (i) Thermal conductivity at 300~K from RTA and iterative LBTE calculations with three-phonon scattering alone and with combined three- and four-phonon scattering.
    In (c--e), circles, squares, and triangles denote DFT, NEP, and MACE, respectively; in (f--h), circles and triangles denote three- and four-phonon processes.
    Transport calculations use a $14^3$ q-point mesh, while the density of states and group velocities are evaluated on a $20^3$ mesh.
    }
    \label{fig:bas_fourphonon}
\end{figure}

Figure~\ref{fig:bas_fourphonon} compares DFT, NEP, and MACE under the same temperature and reciprocal-space settings. All three force models reproduce the characteristic acoustic--optical separation of BAs, although differences remain in the optical frequencies and acoustic group velocities [Fig.~\ref{fig:bas_fourphonon}(b,c)]. These harmonic properties provide a common basis for comparing the higher-order scattering calculations.

The mode-resolved scattering rates reveal differences that are less apparent at the harmonic level. Three-phonon rates span several orders of magnitude and vary strongly from mode to mode, whereas the four-phonon rates show a smoother frequency dependence [Fig.~\ref{fig:bas_fourphonon}(d,e)]. The largest deviations from DFT occur mainly in the optical-frequency range and are more pronounced for three-phonon scattering [Fig.~\ref{fig:bas_fourphonon}(f)]. These differences reflect not only the higher-order force constants themselves, but also the phonon frequencies, eigenvectors, occupation factors, and phase space entering the scattering probabilities. In Fig.~\ref{fig:bas_fourphonon}(f), modes are matched by wave vector and frequency-ordered branch index, with possible branch-index ambiguity near crossings and degeneracies.

The Normal and Umklapp components are shown separately in Fig.~\ref{fig:bas_fourphonon}(g,h). Their relative importance varies with frequency and differs between three- and four-phonon scattering. This distinction is relevant when comparing RTA and iterative transport: Normal processes redistribute nonequilibrium phonon populations without directly relaxing crystal momentum, whereas Umklapp processes contribute to momentum relaxation. Their effect on the heat current is therefore determined by the coupled collision operator rather than by a simple classification into independent resistive lifetimes.

Four-phonon scattering substantially reduces the calculated conductivity for all three force models [Fig.~\ref{fig:bas_fourphonon}(i)]. Within the RTA, the 300~K conductivities decrease from 1450.3, 2040.0, and 1880.0~W\,m$^{-1}$\,K$^{-1}$ for DFT, NEP, and MACE to 894.7, 1110.5, and 971.3~W\,m$^{-1}$\,K$^{-1}$ when four-phonon scattering is included. These changes correspond to reductions of 38.3\%, 45.6\%, and 48.3\%, respectively. The effect is even larger in the iterative LBTE: the three-phonon values of 2518.7, 3020.0, and 3160.1~W\,m$^{-1}$\,K$^{-1}$ decrease to 1291.1, 1354.6, and 1296.4~W\,m$^{-1}$\,K$^{-1}$ when three- and four-phonon collision terms are treated together, corresponding to reductions of 48.7\%, 55.1\%, and 59.0\%.

The difference between RTA and the full LBTE remains substantial after four-phonon scattering is included. The combined-scattering LBTE conductivities exceed their RTA counterparts by 44.3\%, 22.0\%, and 33.5\% for DFT, NEP, and MACE, respectively. The remaining LBTE–RTA difference indicates that off-diagonal collision coupling remains important when four-phonon scattering is included. In the present calculations, the four-phonon collision terms are included directly in the iterative solution rather than introduced only through an additional diagonal scattering rate.

The DFT results also provide a reference for evaluating the MLIP-driven FC4 calculations under identical transport settings. For the three-phonon LBTE, NEP and MACE overestimate the DFT conductivity by 19.9\% and 25.5\%, respectively. After four-phonon scattering is included in the iterative solution, these deviations decrease to 4.9\% and 0.4\%. Despite their similar total conductivities, NEP and MACE exhibit distinct mode-resolved scattering rates and Normal/Umklapp contributions [Fig.~\ref{fig:bas_fourphonon}(c--h)]. The agreement in total conductivity may reflect compensation among differences in phonon velocities, scattering rates, and mode coupling.

\subsection{Wigner population--coherence transport in Cs$_2$AgInCl$_6$}

Cs$_2$AgInCl$_6$ provides a representative case in which the dense vibrational spectrum brings many phonon branches into close spectral proximity. Under these conditions, transport cannot always be described adequately in terms of diagonal phonon populations alone. The Wigner formulation is therefore used to resolve the conventional population contribution, $\kappa_{\mathrm P}$, together with the intermode coherence contribution, $\kappa_{\mathrm C}$~\cite{simoncelli2019,simoncelli2022}.

\begin{figure}[h]
    \centering
    \includegraphics[width=\linewidth]{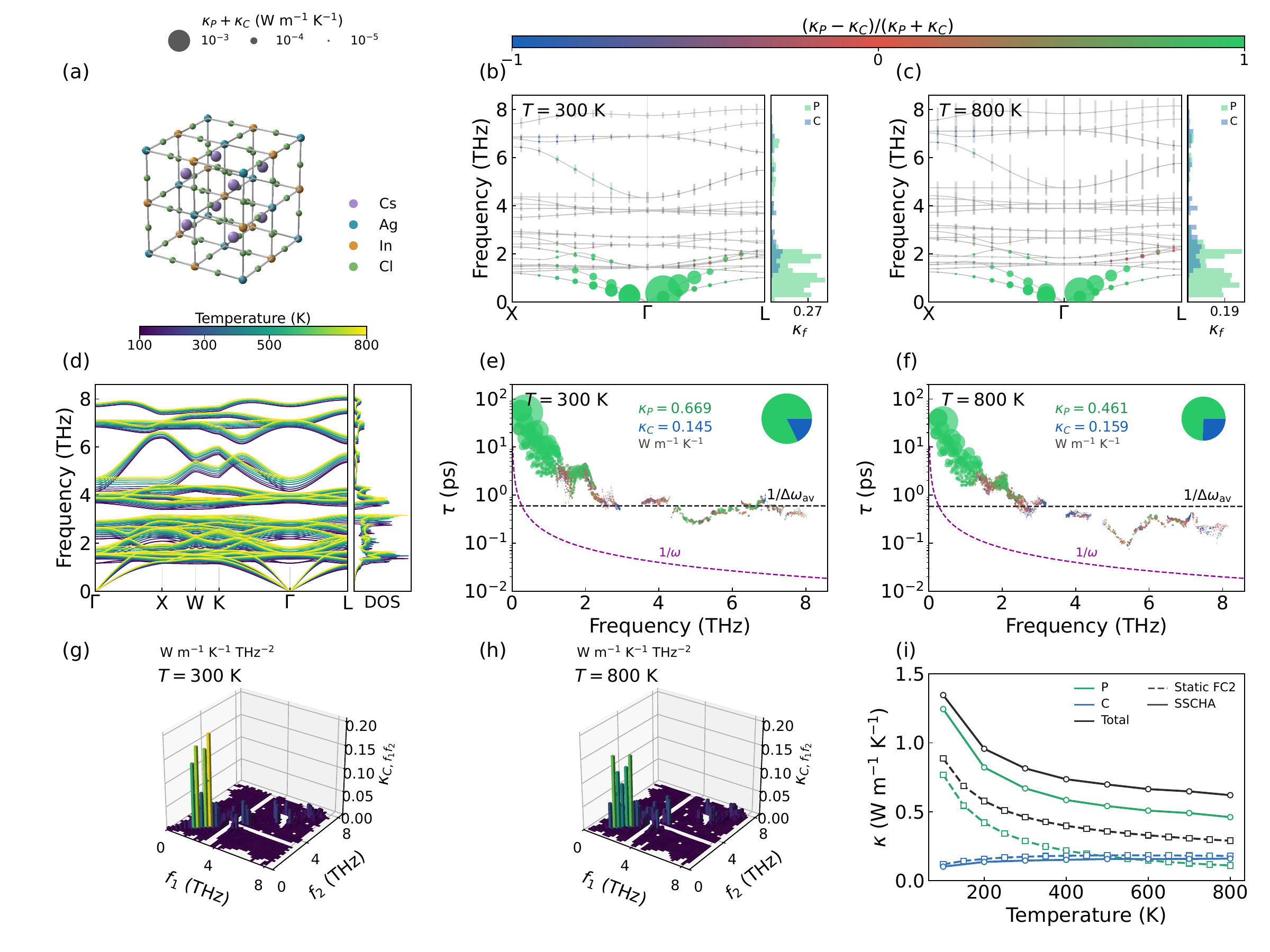}
    \caption{
    Population and coherence transport in Cs$_2$AgInCl$_6$ calculated using the NEP potential.
    (a) Crystal structure.
    (b,c) Phonon dispersions with mode-resolved conductivity contributions at selected temperatures.
    (d) Temperature-dependent auxiliary phonon dispersions and densities of states obtained from the stochastic self-consistent calculation.
    (e,f) Phonon lifetimes as a function of frequency.
    In (b,c,e,f), bubble areas represent the magnitude of the modal conductivity contribution, while colors indicate
    $\eta_\lambda=
    (\kappa_{\mathrm P,\lambda}-\kappa_{\mathrm C,\lambda})/
    (\kappa_{\mathrm P,\lambda}+\kappa_{\mathrm C,\lambda})$.
    (g,h) Frequency-pair-resolved coherence conductivity.
    (i) Temperature dependence of $\kappa_{\mathrm P}$, $\kappa_{\mathrm C}$, and their sum.
    }
    \label{fig:cs2agincl6}
\end{figure}

To visualize the relative character of the mode-resolved transport, Fig.~\ref{fig:cs2agincl6}(b,c,e,f) uses
\begin{equation}
 \eta_\lambda
 =
 \frac{
 \kappa_{\mathrm P,\lambda}
 -
 \kappa_{\mathrm C,\lambda}
 }{
 \kappa_{\mathrm P,\lambda}
 +
 \kappa_{\mathrm C,\lambda}
 } .
 \label{eq:population-coherence-indicator}
\end{equation}
For the nonnegative diagonal contributions considered here,
$\eta_\lambda\rightarrow+1$ denotes predominantly population-like transport, whereas
$\eta_\lambda\rightarrow-1$ denotes predominantly coherence-like transport.
The bubble area represents the magnitude of the modal conductivity contribution, while the color indicates its population–coherence character.

The lifetime distributions in Fig.~\ref{fig:cs2agincl6}(e,f) show that spectral broadening alone is not sufficient to determine whether transport is population or coherence dominated. The coherence term depends simultaneously on the linewidths, the frequency separation between neighboring branches, and the off-diagonal velocity-matrix elements. Modes with comparable lifetimes can therefore exhibit very different values of $\eta_\lambda$. The modal visualization distributes pairwise coherence contributions among individual modes, while the frequency-pair representation retains their original two-mode structure.

The pair-resolved maps in Fig.~\ref{fig:cs2agincl6}(g,h) retain the two frequencies associated with each coherence contribution and therefore expose its interband origin more directly. Contributions concentrated near the diagonal arise from pairs of modes with similar frequencies, whereas weight farther from the diagonal reflects coupling between more widely separated branches. Summation over all frequency pairs recovers the total coherence conductivity, while the distribution identifies the spectral regions that contribute most strongly to $\kappa_{\mathrm C}$.

Finite-temperature renormalization further modifies this balance by changing both the vibrational spectrum and the quantities entering the Wigner transport expression. The auxiliary spectra in Fig.~\ref{fig:cs2agincl6}(d) show the evolution of the branch frequencies with temperature, which changes the separation between neighboring modes. At the same time, changes in the eigenvectors, velocity-matrix elements, and scattering rates affect both the population and coherence channels. The temperature dependence of $\kappa_{\mathrm C}$ therefore reflects the combined evolution of mode spacing, spectral broadening, and interband coupling rather than spectral proximity alone.

The resulting population, coherence, and total conductivities are compared in Fig.~\ref{fig:cs2agincl6}(i). At 300 K, the population and coherence conductivities are 0.669 and 0.145 W m$^{-1}$ K$^{-1}$, respectively. At 800 K, the population contribution decreases to 0.461 W m$^{-1}$ K$^{-1}$, while the coherence contribution reaches 0.159 W m$^{-1}$ K$^{-1}$. The resulting coherence fraction increases from approximately 17.8\% to 25.6\%. This increase reflects both the reduction in population transport and the persistence of interband coherence at elevated temperature.

\subsection{Energy redistribution and thermalization in 3C-SiC}
\label{sec:tdbte}

The time-dependent formulation is applied to a selectively excited phonon population in 3C-SiC. Starting from the Bose--Einstein distribution at $T_0=300$~K, the occupations of all three optical branches are increased by 1\%, corresponding to $\delta=0.01$ in Eq.~\eqref{eq:tdbte-initial}. This perturbation injects $E_{\mathrm{inj}}=7.11\times10^{-5}$~eV per primitive cell into the optical subsystem. No further energy is supplied after initialization. The spatially homogeneous dynamics are followed for 10~ps on a $7^3$ reciprocal-space grid with fixed phonon frequencies and three-phonon collisions, so that the calculation isolates the redistribution of energy in phonon mode space.

The relaxation is examined at three levels: energy transfer between the optical and acoustic subsystems, redistribution among acoustic branches, and relaxation toward a Bose--Einstein distribution within each branch. These processes are characterized by distinct observables and need not occur sequentially. Three-phonon processes are classified as AAA, OAA, OOA, and OOO according to the acoustic (A) or optical (O) character of the three participating modes. These channels involve zero, one, two, and three optical modes, respectively. Each category includes all energetically and momentum-allowed absorption and decay processes with the corresponding mode composition. The labels identify the participating modes and do not, by themselves, specify the direction of net energy transfer.

\begin{figure}[htbp]
    \centering
    \includegraphics[width=\linewidth]{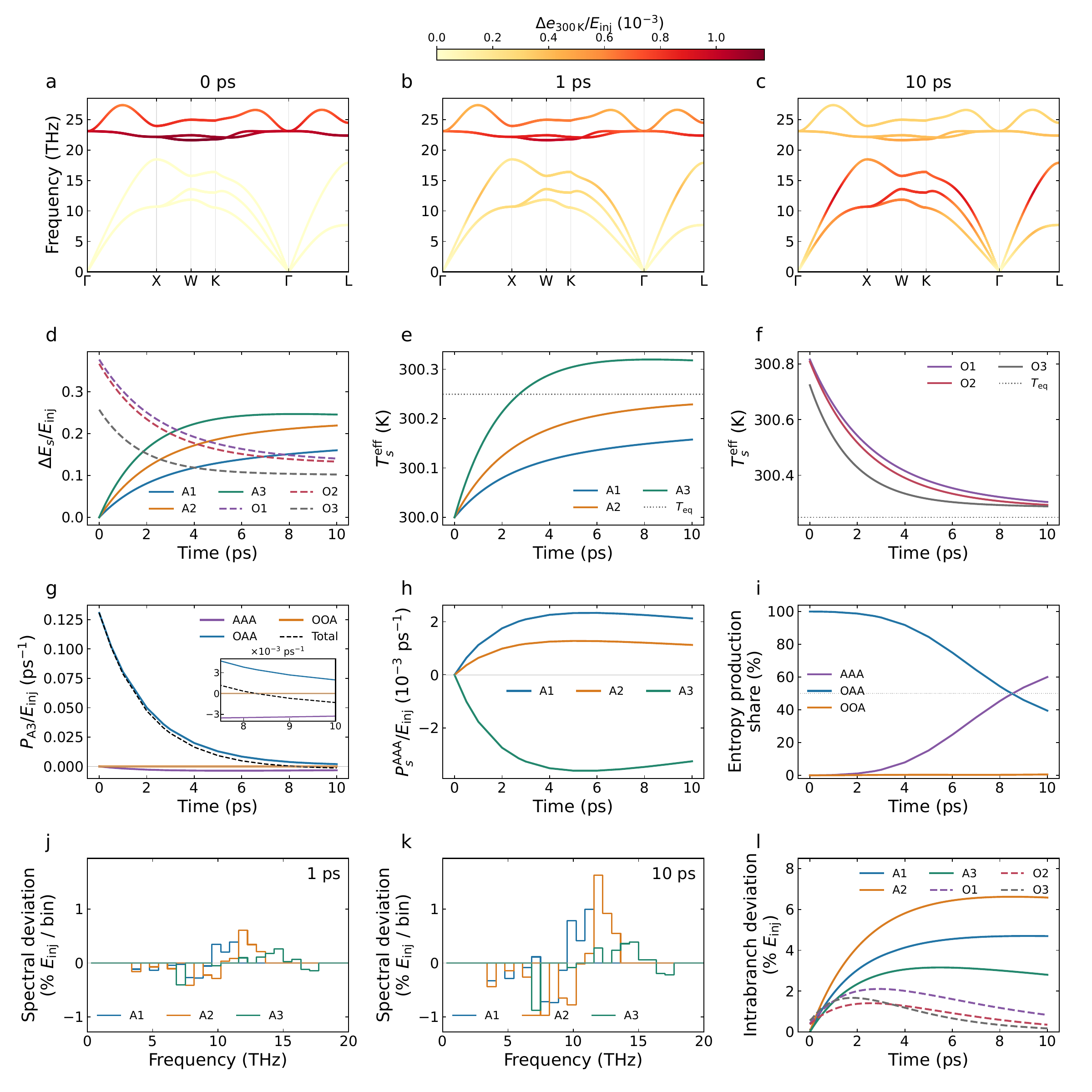}
    \caption{
    Nonequilibrium energy redistribution and thermalization in 3C-SiC following a 1\% increase in the optical-mode occupations at 300~K.
    (a--c) Fixed-frequency phonon dispersions at $t=0$, 1, and 10~ps, colored by excess modal energy normalized by $E_{\mathrm{inj}}$.
    (d) Branch-resolved excess energies; A1--A3 and O1--O3 denote frequency-ordered acoustic and optical branches.
    (e,f) Energy-matched effective temperatures of the acoustic and optical branches; dotted lines indicate the equilibrium temperature determined by energy conservation.
    (g) Net energy-transfer rates into A3 from AAA, OAA, and OOA processes.
    (h) AAA-mediated energy-transfer rates into A1--A3.
    (i) Instantaneous entropy-production fractions associated with the different three-phonon channels.
    (j,k) Frequency-resolved deviations of A1--A3 from their branch-matched Bose--Einstein distributions at 1 and 10~ps.
    (l) Intrabranch nonequilibrium measure $R_s$, expressed as a percentage of $E_{\mathrm{inj}}$.
    }
    \label{fig:sic_time_bte}
\end{figure}

Energy leaves the initially excited optical subsystem rapidly during the first few picoseconds and is progressively transferred to acoustic modes. The acoustic branches contain 22.25\% of $E_{\mathrm{inj}}$ after 1~ps, 55.16\% after 5~ps, and 62.45\% after 10~ps [Fig.~\ref{fig:sic_time_bte}(d)]. Half of the injected energy reaches the acoustic subsystem within approximately 3.75~ps. At the final equilibrium temperature, $T_{\mathrm{eq}}=300.249$~K, acoustic modes would contain 68.40\% of the excess energy, so that about 91\% of the eventual acoustic-energy gain has already occurred by 10~ps. The mode-resolved snapshots in Fig.~\ref{fig:sic_time_bte}(a--c) show that this transfer is strongly nonuniform across the spectrum rather than a uniform increase in acoustic-mode populations.

The individual branches respond on markedly different time scales [Fig.~\ref{fig:sic_time_bte}(e,f)]. Using the first time at which $|T_s^{\mathrm{eff}}-T_{\mathrm{eq}}|$ falls to one half of its initial value as a characteristic response time, the acoustic branches span 0.92--4.70~ps, from A3 to A1, whereas the optical branches lie within the narrower range of 1.39--2.11~ps. The response time of A3 is therefore only about one fifth of that of A1. This broad spread shows that neither the acoustic nor the optical subsystem can be represented by a single relaxation time.

The fastest acoustic branch, A3, also exhibits a pronounced nonmonotonic response. Its effective temperature first reaches $T_{\mathrm{eq}}$ near 2.7~ps, continues to rise, and peaks near 300.320~K at approximately 8.3~ps before decreasing again. The channel decomposition in Fig.~\ref{fig:sic_time_bte}(g) shows that this overshoot results from competition between two distinct energy flows. OAA processes continue to supply energy to A3, while AAA scattering transfers energy from A3 to the other acoustic branches. Near the temperature maximum, the normalized OAA input and AAA output are approximately $+3.41\times10^{-3}$ and $-3.44\times10^{-3}$~ps$^{-1}$, respectively, and almost cancel. By 10~ps, the OAA contribution has fallen to $1.96\times10^{-3}$~ps$^{-1}$, whereas the AAA contribution remains $-3.26\times10^{-3}$~ps$^{-1}$, leaving a net loss of energy from A3 even though optical-to-acoustic transfer is still active.

The corresponding redistribution within the acoustic subsystem is shown in Fig.~\ref{fig:sic_time_bte}(h). AAA processes conserve the total acoustic energy but move energy between acoustic branches: the loss from A3 is accompanied by gains in A1 and A2. These internal energy currents are invisible in the total acoustic-energy balance but are responsible for the distinct branch trajectories. The signed net energy-transfer rates reflect the balance of forward and reverse processes rather than their total scattering activity. The quantities plotted here are signed net powers, so strong forward and reverse scattering may largely cancel even when individual collision events remain frequent.

The dominant source of irreversibility changes as relaxation proceeds. OAA processes account for approximately 99.8\% of the instantaneous entropy production at 1~ps [Fig.~\ref{fig:sic_time_bte}(i)], consistent with the rapid initial transfer of energy from optical to acoustic modes. Their relative contribution decreases as the optical--acoustic imbalance is reduced, while the contribution from AAA processes grows. The two become comparable near 8.5~ps; by 10~ps, AAA contributes about 60\% of the instantaneous entropy production, compared with 39.5\% from OAA and less than 1\% from OOA. The relaxation therefore crosses over from an early stage dominated by inter-subsystem energy transfer to a later stage dominated increasingly by redistribution within the acoustic subsystem.

Branch-energy equilibration does not, however, imply that the occupations within a branch have reached a Bose--Einstein distribution. The mode-resolved deviations defined in Eq.~\eqref{eq:tdbte-shape-energy} provide a direct measure of the remaining intrabranch redistribution. At 10~ps, A1 and A2 retain clear frequency-dependent structure, with deficits below approximately 10~THz and excesses between 10 and 15~THz relative to their branch-matched Bose--Einstein distributions [Fig.~\ref{fig:sic_time_bte}(j,k)]. Because the reference distribution has the same total branch energy, these positive and negative contributions describe redistribution within the branch rather than a net excess or deficit of branch energy.

The residual measure $R_s$ quantifies the magnitude of this intrabranch nonequilibrium [Fig.~\ref{fig:sic_time_bte}(l)]. At 10~ps, $100R_s$ remains 4.69\%, 6.58\%, and 2.80\% for A1, A2, and A3, respectively, whereas the optical-branch values have fallen below 1\%. The acoustic residuals initially increase because the optical-to-acoustic transfer populates different acoustic modes unevenly, and their maxima occur substantially later than those of the optical branches. In particular, A1 and A2 remain strongly nonequilibrium near the end of the trajectory even though most of the eventual acoustic-energy gain has already taken place.

The relaxation therefore proceeds through distinct stages rather than a single characteristic decay. OAA scattering first drives rapid energy transfer from the optical to the acoustic subsystem. AAA scattering then becomes increasingly important in redistributing that energy among the acoustic branches, producing the transient A3 overshoot, while the mode occupations within individual acoustic branches relax on still longer time scales. Inter-subsystem energy transfer, interbranch redistribution, and intrabranch thermalization are thus dynamically distinct processes.

\section{Discussion}

NEP-kappa 2.0 establishes a unified computational framework that connects first-principles calculations, material-specific machine-learning interatomic potentials, and pretrained force models with a broad range of phonon-transport methods. By separating force evaluation from the subsequent lattice-dynamical and transport solvers, the same interatomic description can be carried consistently from harmonic phonons to finite-temperature renormalization, higher-order anharmonic transport, Wigner population--coherence transport, and time-dependent phonon dynamics. This design also enables different force models to be compared under the same structural, numerical, and transport settings, reducing ambiguities introduced by changes in the surrounding workflow.

The calculations presented here demonstrate the main branches of this framework. In 3C-SiC, a material-specific NEP and a pretrained MACE model reproduce the DFT-based phonon transport with 300~K RTA conductivity deviations of approximately $-1.2\%$ and $+8.4\%$, respectively. Finite-temperature calculations further show that thermal expansion and intrinsic anharmonicity can act either competitively or cooperatively depending on the phonon mode, and that their combined effect on transport cannot be inferred from a uniform shift of the phonon spectrum. In BAs, the framework is extended to FC4 and combined three- and four-phonon transport, where four-phonon scattering lowers the 300~K RTA conductivity by 38.3--48.3\% for the three force models considered. The comparison between RTA and iterative solutions also shows that higher-order scattering does not remove the importance of population redistribution beyond the diagonal relaxation-time picture. For Cs$_2$AgInCl$_6$, the Wigner treatment resolves the coexistence of population and coherence transport in a dense vibrational spectrum and relates the coherence contribution to mode spacing, linewidths, and off-diagonal velocity-matrix elements.

The time-dependent calculation reveals a multistage thermalization process that is not accessible from steady-state conductivities or equilibrium scattering rates alone. Following selective excitation of the optical branches in 3C-SiC, half of the injected energy is transferred to acoustic modes within approximately 3.75~ps. OAA processes dominate this early optical-to-acoustic transfer, whereas AAA scattering progressively redistributes energy among the acoustic branches. The resulting competition produces a transient overshoot of the A3 branch and branch response times spanning approximately 0.92--4.70~ps. At later times, AAA processes become increasingly important while substantial non-Bose occupation patterns remain within the acoustic branches. These results distinguish energy transfer between phonon subsystems, redistribution among branches, and intrabranch thermalization as separate stages of the relaxation process.

\section{Methods}

\subsection{Force constants and harmonic lattice dynamics}

The lattice potential energy is expanded around a reference structure in powers of the atomic displacements $u_i$, where the composite index $i$ identifies the unit cell, atom, and Cartesian direction~\cite{born1954,srivastava2023}:
\begin{equation}
 U=U_0+\frac{1}{2!}\sum_{ij}\Phi^{(2)}_{ij}u_i u_j
 +\frac{1}{3!}\sum_{ijk}\Phi^{(3)}_{ijk}u_i u_j u_k
 +\frac{1}{4!}\sum_{ijkl}\Phi^{(4)}_{ijkl}u_i u_j u_k u_l+\cdots .
 \label{eq:ifc-expansion}
\end{equation}
The linear term vanishes at mechanical equilibrium. The second-order force constants (FC2) determine the harmonic lattice dynamics, whereas FC3 and FC4 describe cubic and quartic anharmonic interactions, respectively. In NEP-kappa 2.0, these force constants can be extracted from displacement--force data generated with either density-functional theory (DFT) or machine-learning interatomic potentials (MLIPs).

Using the Bloch convention adopted here, the mass-weighted dynamical matrix is
\begin{equation}
 D_{\kappa\alpha,\kappa'\beta}(\mathbf q)
 =\frac{1}{\sqrt{M_\kappa M_{\kappa'}}}
 \sum_l
 \Phi^{(2)}_{0\kappa\alpha,l\kappa'\beta}
 e^{i\mathbf q\cdot
 (\mathbf R_l+\boldsymbol\tau_{\kappa'}-\boldsymbol\tau_\kappa)} ,
 \label{eq:dynamical-matrix}
\end{equation}
where $M_\kappa$ and $\boldsymbol\tau_\kappa$ are the mass and basis position of atom $\kappa$, respectively. The phonon frequencies and eigenvectors are obtained from
\begin{equation}
 D(\mathbf q)\mathbf e_{\mathbf q s}
 =\omega_{\mathbf q s}^{2}\mathbf e_{\mathbf q s}.
 \label{eq:phonon-eigenvalue}
\end{equation}
We denote a phonon mode by $\lambda\equiv(\mathbf q,s)$, and use $\omega_\lambda$ for its angular frequency. Frequencies reported in THz are expressed as
$\nu_\lambda=\omega_\lambda/(2\pi)$.

The group velocity of mode $\lambda$ is
\begin{equation}
 \mathbf v_\lambda=\nabla_{\mathbf q}\omega_\lambda ,
 \label{eq:group-velocity}
\end{equation}
and the phonon density of states is written as
\begin{equation}
 g(\omega)
 =\frac{1}{N_q}
 \sum_{\lambda}
 \delta(\omega-\omega_\lambda),
 \label{eq:phonon-dos}
\end{equation}
where $N_q$ is the number of sampled wave vectors in the Brillouin zone.

The harmonic heat capacity of mode $\lambda$ is
\begin{equation}
 C_\lambda(T)
 =
 k_{\mathrm B}x_\lambda^2
 \frac{e^{x_\lambda}}
 {(e^{x_\lambda}-1)^2},
 \qquad
 x_\lambda=
 \frac{\hbar\omega_\lambda}{k_{\mathrm B}T},
 \label{eq:mode-heat-capacity}
\end{equation}
which gives the volumetric constant-volume heat capacity
\begin{equation}
 C_V^{\mathrm{vol}}(T)
 =
 \frac{1}{V N_q}
 \sum_{\lambda}
 C_\lambda(T),
 \label{eq:volumetric-heat-capacity}
\end{equation}
where $V$ is the primitive-cell volume.

\subsection{Finite-temperature phonons: QHA and SSCHA}

Finite-temperature changes in the phonon spectrum are treated by separating the effect of thermal expansion from intrinsic anharmonic renormalization. The former is described within the quasi-harmonic approximation (QHA), while the latter is evaluated using the stochastic self-consistent harmonic approximation (SSCHA)~\cite{togo2010_qha,errea2014,monacelli2021}.

Within QHA, the phonons remain harmonic at each fixed volume, but their frequencies vary with the thermally induced change in lattice volume. For a given volume $V$, the vibrational Helmholtz free energy per primitive cell is
\begin{equation}
 F_{\mathrm{vib}}(V,T)
 =
 \frac{1}{N_q}
 \sum_{\mathbf q,s}
 \left[
 \frac{\hbar\omega_{\mathbf q s}(V)}{2}
 +k_{\mathrm B}T
 \ln\left(
 1-e^{-\hbar\omega_{\mathbf q s}(V)/(k_{\mathrm B}T)}
 \right)
 \right].
 \label{eq:qha-free-energy}
\end{equation}
At an external pressure $P$, the equilibrium volume follows from minimization of the Gibbs free energy,
\begin{equation}
 V_{\mathrm{QHA}}(T,P)
 =
 \underset{V}{\operatorname{arg\,min}}
 \left[
 U_0(V)+F_{\mathrm{vib}}(V,T)+PV
 \right],
 \label{eq:qha-volume}
\end{equation}
where $U_0(V)$ is the static lattice energy~\cite{togo2010_qha}. The volumetric thermal-expansion coefficient is then
\begin{equation}
 \alpha_V(T)
 =
 \frac{1}{V_{\mathrm{QHA}}}
 \left(
 \frac{\partial V_{\mathrm{QHA}}}{\partial T}
 \right)_P .
 \label{eq:thermal-expansion}
\end{equation}
The sensitivity of an individual phonon mode to volume is characterized by the mode Gr\"uneisen parameter,
\begin{equation}
 \gamma_\lambda
 =
 -\frac{\partial\ln\omega_\lambda}
 {\partial\ln V}.
 \label{eq:mode-gruneisen}
\end{equation}
Thus, QHA incorporates temperature dependence through the equilibrium volume while retaining a harmonic description at each volume.

Intrinsic anharmonic effects at a specified volume are treated with SSCHA, which determines an optimal auxiliary harmonic Hamiltonian by variational minimization of the anharmonic free energy~\cite{errea2014,monacelli2021}. For a trial harmonic Hamiltonian characterized by force constants $\Phi$ and thermal density matrix $\rho_\Phi$, the variational functional can be written as
\begin{equation}
 \mathcal F[\Phi]
 =
 F_\Phi^{\mathrm{harm}}
 +
 \left\langle
 U-U_\Phi^{\mathrm{harm}}
 \right\rangle_{\rho_\Phi}.
 \label{eq:sscha-functional}
\end{equation}
The required ensemble averages are evaluated by stochastic sampling of configurations drawn from the trial harmonic density matrix. Their energies and forces are calculated with the selected force engine, and the trial harmonic force constants are updated until the variational solution is converged. In the present calculations, the cell volume is fixed during this procedure. The resulting effective second-order force constants are denoted by
\begin{equation}
 \Phi_{\mathrm{eff}}^{(2)}(T,V).
 \label{eq:sscha-effective-fc2}
\end{equation}

The eigenvalues of the auxiliary dynamical matrix are used as effective finite-temperature phonon frequencies in the subsequent lattice-dynamical and transport calculations. These auxiliary frequencies are distinct from those obtained from the SSCHA free-energy Hessian or the full dynamical spectral function~\cite{bianco2017,monacelli2021}. The latter require additional response or self-energy calculations beyond the variational auxiliary Hamiltonian.

To account for thermal expansion and intrinsic anharmonicity within the same structural reference, NEP-kappa 2.0 employs a sequential QHA+SSCHA procedure. At each temperature, QHA first determines the equilibrium volume, after which the SSCHA calculation is carried out at that volume:
\begin{equation}
 \Phi_{\mathrm{eff}}^{(2)}
 \left[T,V_{\mathrm{QHA}}(T)\right].
 \label{eq:qha-sscha-fc2}
\end{equation}
This sequential procedure evaluates intrinsic anharmonic renormalization at the temperature-dependent QHA equilibrium volume.

For SSCHA-based transport calculations, the static FC2 is replaced by the temperature-dependent effective force constants $\Phi_{\mathrm{eff}}^{(2)}(T,V)$, while FC3 and, when applicable, FC4 are calculated at the corresponding reference structure without explicit finite-temperature renormalization. In the combined QHA+SSCHA treatment, the higher-order force constants are recalculated at each QHA equilibrium volume $V_{\mathrm{QHA}}(T)$. The effective phonon frequencies and eigenvectors are subsequently used to evaluate group velocities, scattering phase space, and anharmonic interaction matrix elements.

\subsection{Three- and four-phonon scattering and population transport}

Anharmonic phonon scattering is determined by the higher-order force constants introduced above. FC3 gives the cubic interaction matrix elements responsible for three-phonon processes, whereas FC4 generates four-phonon scattering~\cite{srivastava2023,feng2016,feng2018}. The allowed processes are constrained by energy and crystal-momentum conservation. Three-phonon scattering consists of absorption and decay processes, while the four-phonon phase space additionally contains one-to-three and two-to-two channels and their reverse processes. The corresponding transition rates depend on both the available phase space and the anharmonic interaction matrix elements, together with the Bose--Einstein occupations of the participating modes~\cite{feng2014prediction}.

For a diagonal relaxation-time treatment, the total scattering rate of mode $\lambda$ is written as
\begin{equation}
 \Gamma_\lambda
 =
 \Gamma_{3,\lambda}
 +
 \Gamma_{4,\lambda},
 \qquad
 \tau_\lambda=\Gamma_\lambda^{-1},
 \label{eq:total-scattering-rate}
\end{equation}
when three- and four-phonon processes are both included. Additional independent scattering mechanisms can be incorporated in the same manner when required~\cite{klemens1959,tamura1983}. This additive form refers to the diagonal relaxation rates; in an iterative solution of the BTE, scattering enters through the full collision operator rather than through a simple sum of mode lifetimes.

Within the relaxation-time approximation (RTA), each phonon mode relaxes independently, giving the population contribution to the thermal-conductivity tensor as~\cite{broido2007,li2014_shengbte}
\begin{equation}
 \kappa_{\mathrm P}^{\alpha\beta}
 =
 \frac{1}{V N_q}
 \sum_{\lambda}
 C_\lambda
 v_\lambda^\alpha
 v_\lambda^\beta
 \tau_\lambda ,
 \label{eq:kappa-rta}
\end{equation}
where $C_\lambda$, $\mathbf v_\lambda$, and $\tau_\lambda$ are the modal heat capacity, group velocity, and relaxation time, respectively.

Beyond the RTA, the linearized BTE can be solved by retaining the coupling between nonequilibrium phonon populations. Two numerical approaches are used in NEP-kappa 2.0. The direct method of Chaput constructs the collision matrix and obtains the transport solution through its diagonalization and pseudoinverse~\cite{chaput2013}. Alternatively, the coupled BTE can be solved iteratively, as implemented in ShengBTE~\cite{omini1995,li2014_shengbte}. In either case, the conductivity can be expressed in terms of the generalized mean free displacement $\boldsymbol{\ell}_\lambda$,
\begin{equation}
 \kappa_{\mathrm P}^{\alpha\beta}
 =
 \frac{1}{V N_q}
 \sum_{\lambda}
 C_\lambda
 v_\lambda^\alpha
 \ell_\lambda^\beta .
 \label{eq:kappa-iterative}
\end{equation}
Here, $\boldsymbol{\ell}_\lambda$ is obtained from the solution of the full linearized collision problem and therefore contains the redistribution of the nonequilibrium phonon population among coupled modes. In contrast to the RTA, it is not determined solely by the lifetime of mode $\lambda$, but also by scattering into and out of other modes through the off-diagonal elements of the collision operator. Neglecting these mode-coupling terms reduces the solution to the RTA limit,
$\boldsymbol{\ell}_\lambda=\mathbf v_\lambda\tau_\lambda$.

For calculations including four-phonon scattering, the present full BTE treatment follows the iterative scheme implemented in FourPhonon~\cite{han2022_fourphonon,han2023graphene,li2024convergent,jain2024fourphonon}. Three- and four-phonon collision terms are then included self-consistently in the iterative solution. This differs from the commonly used approximation in which the three-phonon BTE is solved beyond the RTA while four-phonon scattering enters only through an additional diagonal relaxation rate.

\subsection{Wigner population and coherence transport}

The Wigner transport equation extends the Peierls--Boltzmann description by retaining both the diagonal and off-diagonal components of the phonon distribution matrix~\cite{simoncelli2019,simoncelli2022}. At a given wave vector $\mathbf q$, the diagonal elements describe phonon populations and recover the conventional Boltzmann transport problem, whereas the off-diagonal elements describe coherences between different vibrational branches. The thermal conductivity is correspondingly written as
$\kappa_{\mathrm{tot}}=\kappa_{\mathrm P}+\kappa_{\mathrm C}$,
where $\kappa_{\mathrm P}$ is the population contribution introduced in the preceding section and $\kappa_{\mathrm C}$ arises from interband coherences.

Unlike the conventional BTE, which involves only the diagonal group velocities, the coherence term requires the full velocity matrix. In the phonon eigenmode basis,
\begin{equation}
 \mathsf v_{ss'}^\alpha(\mathbf q)
 =
 \sum_{\kappa\beta,\kappa'\beta'}
 e_{\mathbf q s,\kappa\beta}^{*}
 \left[
 \frac{\partial \sqrt{D(\mathbf q)}}
 {\partial q_\alpha}
 \right]_{\kappa\beta,\kappa'\beta'}
 e_{\mathbf q s',\kappa'\beta'} ,
 \label{eq:wigner-velocity}
\end{equation}
where $D(\mathbf q)$ is the dynamical matrix and
$e_{\mathbf q s}$ is the phonon eigenvector~\cite{simoncelli2022}.
The diagonal element
$\mathsf v_{ss}^{\alpha}(\mathbf q)=v_{\mathbf q s}^{\alpha}$
reduces to the usual phonon group velocity, while the off-diagonal elements couple different branches at the same wave vector. The velocity matrix is evaluated using the smooth Bloch-phase convention, which gives a Wigner conductivity that is independent of the particular choice of crystal unit cell and remains size consistent~\cite{simoncelli2022}.

The strength of the coherence contribution depends on the interplay among interband frequency separation, phonon broadening, and the off-diagonal velocity matrix elements. Following the linewidth convention of Ref.~\cite{simoncelli2022}, $\Gamma_{\mathbf q s}=\tau_{\mathbf q s}^{-1}$, and for a pair of modes we define
$\bar{\Gamma}_{ss'}(\mathbf q)
=[\Gamma_{\mathbf q s}+\Gamma_{\mathbf q s'}]/2$.
The coherence contribution to the thermal-conductivity tensor is then
\begin{equation}
\begin{split}
 \kappa_{\mathrm C}^{\alpha\beta}
 =
 \frac{1}{V N_q}
 \sum_{\mathbf q}
 \sum_{s\ne s'}
 &\frac{\omega_{\mathbf q s}+\omega_{\mathbf q s'}}{4}
 \left(
 \frac{C_{\mathbf q s}}{\omega_{\mathbf q s}}
 +
 \frac{C_{\mathbf q s'}}{\omega_{\mathbf q s'}}
 \right)\\
 &\times
 \operatorname{Re}
 \left[
 \mathsf v_{ss'}^\alpha(\mathbf q)
 \mathsf v_{s's}^\beta(\mathbf q)
 \right]
 \frac{
 \bar{\Gamma}_{ss'}(\mathbf q)
 }{
 \left(
 \omega_{\mathbf q s}
 -
 \omega_{\mathbf q s'}
 \right)^2
 +
 \bar{\Gamma}_{ss'}^2(\mathbf q)
 } ,
 \label{eq:wigner-coherence}
\end{split}
\end{equation}
where $C_{\mathbf q s}$ is the modal heat capacity and $V$ is the primitive-cell volume~\cite{simoncelli2019,simoncelli2022}.

Equation~\eqref{eq:wigner-coherence} shows that coherence is enhanced when the frequency separation between two branches becomes comparable to their spectral broadening, provided that the corresponding off-diagonal velocity matrix element is appreciable. For well-separated phonon bands, the Lorentzian factor suppresses the off-diagonal term and transport approaches the usual Peierls--Boltzmann limit. Dense vibrational spectra, by contrast, provide many near-degenerate mode pairs for which the coherence contribution can become significant~\cite{simoncelli2019,simoncelli2022}.

For the frequency-pair analysis used below, each off-diagonal contribution in Eq.~\eqref{eq:wigner-coherence} is assigned to the pair
$(\omega_{\mathbf q s},\omega_{\mathbf q s'})$.
The resulting two-frequency distribution is constructed so that summation over all frequency pairs recovers the corresponding coherence conductivity $\kappa_{\mathrm C}$.

\subsection{Time-dependent phonon population dynamics}

The time-dependent branch describes the redistribution of energy among phonon modes following a prescribed nonequilibrium excitation. For the spatially homogeneous systems considered here, the occupation of mode $\lambda$ evolves according to the phonon Boltzmann equation~\cite{spohn2006,ono2017,caruso2022,yao2025_rtbte},
\begin{equation}
\frac{d n_\lambda(t)}{dt}
=
\mathcal I_\lambda[\mathbf n(t)],
\label{eq:tdbte}
\end{equation}
where $\mathcal I_\lambda$ is the phonon--phonon collision integral evaluated from the instantaneous mode occupations. The calculations presented in this work include three-phonon collisions. Phonon frequencies are kept fixed during the evolution, and neither spatial drift nor external energy input is included after initialization. The resulting dynamics therefore isolate the redistribution of energy in phonon mode space.

Relative to an equilibrium reference state at temperature $T_0$, the excess energy associated with mode $\lambda=(\mathbf q,s)$ is
\begin{equation}
\Delta\varepsilon_{\mathbf q s}(t)
=
w_{\mathbf q}\hbar\omega_{\mathbf q s}
\left[
n_{\mathbf q s}(t)
-
n_{\mathbf q s}^{\mathrm{BE}}(T_0)
\right],
\label{eq:tdbte-modal-energy}
\end{equation}
where $w_{\mathbf q}$ is the normalized reciprocal-space weight, with $w_{\mathbf q}=1/N_q$ for a uniform full grid. The excess energy carried by branch $s$ is then
\begin{equation}
\Delta E_s(t)
=
\sum_{\mathbf q}
\Delta\varepsilon_{\mathbf q s}(t).
\label{eq:tdbte-branch-energy}
\end{equation}
Frequency-resolved energy distributions are obtained by accumulating the same modal contributions in frequency bins.

A selective nonequilibrium state is introduced by modifying the occupations of a chosen subset of phonon modes,
\begin{equation}
n_\lambda(0)
=
n_\lambda^{\mathrm{BE}}(T_0)
\left[
1+\delta\,\chi_{\mathrm{exc}}(\lambda)
\right],
\label{eq:tdbte-initial}
\end{equation}
where $\chi_{\mathrm{exc}}(\lambda)$ selects the excited modes and $\delta$ specifies the excitation amplitude. The corresponding injected energy is
$E_{\mathrm{inj}}=\sum_s\Delta E_s(0)$.
For the closed collision dynamics considered here, three-phonon scattering conserves the total vibrational energy,
\begin{equation}
\sum_\lambda
w_{\mathbf q}
\hbar\omega_\lambda
\mathcal I_\lambda[\mathbf n(t)]
=
0,
\label{eq:tdbte-energy-conservation}
\end{equation}
so that energy lost by one subset of modes must be transferred to another.

To identify the microscopic pathways of energy redistribution, the collision integral is decomposed into selected classes of three-phonon processes,
\begin{equation}
\mathcal I_\lambda
=
\sum_c
\mathcal I_\lambda^{(c)},
\label{eq:tdbte-channel-decomposition}
\end{equation}
where $c$ labels a class of scattering events. The net rate at which channel $c$ transfers energy into a subset of modes $\mathcal S$ is
\begin{equation}
\dot E_{\mathcal S}^{(c)}(t)
=
\sum_{\lambda\in\mathcal S}
w_{\mathbf q}\hbar\omega_\lambda
\mathcal I_\lambda^{(c)}[\mathbf n(t)].
\label{eq:tdbte-channel-energy-transfer}
\end{equation}
A positive value denotes a net energy flow into $\mathcal S$, whereas a negative value denotes a net loss. This quantity is distinct from a scattering rate: it retains the direction of energy flow and reflects the imbalance between forward and reverse processes at the instantaneous nonequilibrium occupations.

The same decomposition can be used to analyze the instantaneous production of Bose entropy, consistent with the H-theorem for the phonon Boltzmann equation~\cite{spohn2006}. The contribution associated with channel $c$ is
\begin{equation}
\dot S^{(c)}(t)
=
k_{\mathrm B}
\sum_\lambda
w_{\mathbf q}
\ln\left[
\frac{1+n_\lambda(t)}
{n_\lambda(t)}
\right]
\mathcal I_\lambda^{(c)}[\mathbf n(t)].
\label{eq:tdbte-channel-entropy}
\end{equation}
The normalized contribution of channel $c$ is defined as
\begin{equation}
f_S^{(c)}(t)
=
\frac{\dot S^{(c)}(t)}
{\sum_{c'}\dot S^{(c')}(t)}.
\label{eq:tdbte-entropy-share}
\end{equation}
This quantity identifies which classes of phonon collisions contribute most strongly to the instantaneous approach toward equilibrium.

The energetic state of branch $s$ is characterized by an effective temperature $T_s^{\mathrm{eff}}(t)$ obtained by matching its instantaneous energy to that of a Bose--Einstein distribution,
\begin{equation}
\sum_{\mathbf q}
w_{\mathbf q}\hbar\omega_{\mathbf q s}
n_{\mathbf q s}(t)
=
\sum_{\mathbf q}
w_{\mathbf q}\hbar\omega_{\mathbf q s}
n^{\mathrm{BE}}
\left(
\omega_{\mathbf q s},
T_s^{\mathrm{eff}}(t)
\right).
\label{eq:branch-effective-temperature}
\end{equation}
The effective temperature therefore characterizes the total energy of a branch, but does not imply that the occupations within that branch are themselves thermal.

To quantify deviations from the branch-matched Bose--Einstein distribution, we define the signed modal-energy deviation
\begin{equation}
\delta e_{\mathbf q s}(t)
=
w_{\mathbf q}\hbar\omega_{\mathbf q s}
\left[
n_{\mathbf q s}(t)
-
n^{\mathrm{BE}}
\left(
\omega_{\mathbf q s},
T_s^{\mathrm{eff}}(t)
\right)
\right].
\label{eq:tdbte-shape-energy}
\end{equation}
Energy matching ensures that
$\sum_{\mathbf q}\delta e_{\mathbf q s}(t)=0$
for each branch, even when the individual mode occupations remain far from a Bose--Einstein distribution. The magnitude of this intrabranch nonequilibrium is measured by
\begin{equation}
R_s(t)
=
\frac{
\sum_{\mathbf q}
\left|
\delta e_{\mathbf q s}(t)
\right|
}{
E_{\mathrm{inj}}
}.
\label{eq:intrabranch-nonequilibrium}
\end{equation}
The quantities $T_s^{\mathrm{eff}}$ and $R_s$ therefore describe complementary aspects of relaxation: the former measures branch-energy equilibration, whereas the latter measures the remaining redistribution required within the branch.

\section*{Declaration of competing interest}
The authors have no conflicts to disclose.

\section*{Author contributions}
S.L. and B.L. contributed equally to this work. S.L. conceived the computational framework, developed the methodology, performed simulations and data analysis, prepared the figures, and drafted the manuscript. B.L. contributed to methodology development, simulations, data analysis, visualization, and manuscript preparation and revision. F.Y. contributed to methodology development, software implementation, and computational investigations. G.W. contributed to methodology development, computational investigations, and data analysis. Y.D. contributed to software development and computational investigations. Y.Q. contributed to methodology development and manuscript revision. A.V.K. and Q.X. contributed to validation and manuscript review and editing. A.A.B. supervised the research and revised the manuscript. K.X. contributed to conceptualization, project administration, supervision, and manuscript revision. G.Z. contributed to conceptualization, computational investigations, validation, supervision, and manuscript revision. All authors reviewed and approved the final manuscript.

\section*{Data availability}
The data supporting the findings of this study, including interatomic potentials, atomic structures, force constants, and thermal transport results, are available in the NEP-kappa-2.0 dataset at \url{https://github.com/lyushisyan/Research-data/tree/main/2026-nep-kappa-2.0}.

\section*{Code availability}
The NEP-kappa source code used in this study is publicly available at:
\url{https://github.com/lyushisyan/NEP-kappa}.

\section*{Acknowledgements}
This work was supported by the China Scholarship Council through scholarships awarded to S.L. (No. 202308090243), F.Y. (No. 202408090635), Y.D. (No. 202608090077) and Q.X. (202606290026). 
The authors thank Matvey I. Chikin and Vladimir I. Khvesyuk
for helpful discussions.

\bibliographystyle{elsarticle-num}
\bibliography{my-refs}

@article{togo2015_phonon,
    author  = {Togo, Atsushi and Tanaka, Isao},
    title   = {First principles phonon calculations in materials science},
    journal = {Scr. Mater.},
    volume  = {108},
    pages   = {1--5},
    year    = {2015},
    doi     = {10.1016/j.scriptamat.2015.07.021},
}

@article{togo2023_phonopy,
    author  = {Togo, Atsushi},
    title   = {First-principles Phonon Calculations with Phonopy and Phono3py},
    journal = {J. Phys. Soc. Jpn.},
    volume  = {92},
    number  = {1},
    pages   = {012001},
    year    = {2023},
    doi     = {10.7566/JPSJ.92.012001},
}

@article{omini1995,
    author  = {Omini, M. and Sparavigna, A.},
    title   = {An iterative approach to the phonon Boltzmann equation in the theory of thermal conductivity},
    journal = {Physica B},
    volume  = {212},
    number  = {2},
    pages   = {101--112},
    year    = {1995},
    doi     = {10.1016/0921-4526(95)00016-3},
}

@article{broido2007,
    author  = {Broido, D. A. and Malorny, M. and Birner, G. and Mingo, Natalio and Stewart, D. A.},
    title   = {Intrinsic lattice thermal conductivity of semiconductors from first principles},
    journal = {Appl. Phys. Lett.},
    volume  = {91},
    number  = {23},
    pages   = {231922},
    year    = {2007},
    doi     = {10.1063/1.2822891},
}

@article{lindsay2008,
    author  = {Lindsay, Lucas and Broido, D. A.},
    title   = {Three-phonon phase space and lattice thermal conductivity in semiconductors},
    journal = {J. Phys.: Condens. Matter},
    volume  = {20},
    number  = {16},
    pages   = {165209},
    year    = {2008},
    doi     = {10.1088/0953-8984/20/16/165209},
}

@article{esfarjani2011,
    author  = {Esfarjani, Keivan and Chen, Gang and Stokes, Harold T.},
    title   = {Heat transport in silicon from first-principles calculations},
    journal = {Phys. Rev. B},
    volume  = {84},
    number  = {8},
    pages   = {085204},
    year    = {2011},
    doi     = {10.1103/PhysRevB.84.085204},
}

@article{chaput2013,
    author  = {Chaput, Laurent},
    title   = {Direct Solution to the Linearized Phonon Boltzmann Equation},
    journal = {Phys. Rev. Lett.},
    volume  = {110},
    number  = {26},
    pages   = {265506},
    year    = {2013},
    doi     = {10.1103/PhysRevLett.110.265506},
}

@article{li2014_shengbte,
    author  = {Li, Wu and Carrete, Jes{\'u}s and Katcho, Nebil A. and Mingo, Natalio},
    title   = {{ShengBTE}: A solver of the Boltzmann transport equation for phonons},
    journal = {Comput. Phys. Commun.},
    volume  = {185},
    number  = {6},
    pages   = {1747--1758},
    year    = {2014},
    doi     = {10.1016/j.cpc.2014.02.015},
}

@article{togo2015_lifetime,
    author  = {Togo, Atsushi and Chaput, Laurent and Tanaka, Isao},
    title   = {Distributions of phonon lifetimes in Brillouin zones},
    journal = {Phys. Rev. B},
    volume  = {91},
    number  = {9},
    pages   = {094306},
    year    = {2015},
    doi     = {10.1103/PhysRevB.91.094306},
}

@article{feng2016,
    author  = {Feng, Tianli and Ruan, Xiulin},
    title   = {Quantum mechanical prediction of four-phonon scattering rates and reduced thermal conductivity of solids},
    journal = {Phys. Rev. B},
    volume  = {93},
    number  = {4},
    pages   = {045202},
    year    = {2016},
    doi     = {10.1103/PhysRevB.93.045202},
}

@article{feng2018,
    author  = {Feng, Tianli and Ruan, Xiulin},
    title   = {Four-phonon scattering reduces intrinsic thermal conductivity of graphene and the contributions from flexural phonons},
    journal = {Phys. Rev. B},
    volume  = {97},
    number  = {4},
    pages   = {045202},
    year    = {2018},
    doi     = {10.1103/PhysRevB.97.045202},
}

@article{yang2019_fourphonon,
    author  = {Yang, Xiaolong and Feng, Tianli and Li, Ju and Ruan, Xiulin},
    title   = {Stronger role of four-phonon scattering than three-phonon scattering in thermal conductivity of {III-V} semiconductors at room temperature},
    journal = {Phys. Rev. B},
    volume  = {100},
    number  = {24},
    pages   = {245203},
    year    = {2019},
    doi     = {10.1103/PhysRevB.100.245203},
}

@article{han2022_fourphonon,
    author  = {Han, Zherui and Yang, Xiaolong and Li, Wu and Feng, Tianli and Ruan, Xiulin},
    title   = {{FourPhonon}: An extension module to {ShengBTE} for computing four-phonon scattering rates and thermal conductivity},
    journal = {Comput. Phys. Commun.},
    volume  = {270},
    pages   = {108179},
    year    = {2022},
    doi     = {10.1016/j.cpc.2021.108179},
}

@article{togo2010_qha,
    author  = {Togo, Atsushi and Chaput, Laurent and Tanaka, Isao and Hug, Gilles},
    title   = {First-principles phonon calculations of thermal expansion in {Ti3SiC2}, {Ti3AlC2}, and {Ti3GeC2}},
    journal = {Phys. Rev. B},
    volume  = {81},
    number  = {17},
    pages   = {174301},
    year    = {2010},
    doi     = {10.1103/PhysRevB.81.174301},
}

@article{hellman2013_tdep,
    author  = {Hellman, Olle and Steneteg, Peter and Abrikosov, I. A. and Simak, S. I.},
    title   = {Temperature dependent effective potential method for accurate free energy calculations of solids},
    journal = {Phys. Rev. B},
    volume  = {87},
    number  = {10},
    pages   = {104111},
    year    = {2013},
    doi     = {10.1103/PhysRevB.87.104111},
}

@article{hellman2013_fc3,
    author  = {Hellman, Olle and Abrikosov, I. A.},
    title   = {Temperature-dependent effective third-order interatomic force constants from first principles},
    journal = {Phys. Rev. B},
    volume  = {88},
    number  = {14},
    pages   = {144301},
    year    = {2013},
    doi     = {10.1103/PhysRevB.88.144301},
}

@article{errea2014,
    author  = {Errea, Ion and Calandra, Matteo and Mauri, Francesco},
    title   = {Anharmonic free energies and phonon dispersions from the stochastic self-consistent harmonic approximation: Application to platinum and palladium hydrides},
    journal = {Phys. Rev. B},
    volume  = {89},
    number  = {6},
    pages   = {064302},
    year    = {2014},
    doi     = {10.1103/PhysRevB.89.064302},
}

@article{tadano2015,
    author  = {Tadano, Terumasa and Tsuneyuki, Shinji},
    title   = {Self-consistent phonon calculations of lattice dynamical properties in cubic {SrTiO3} with first-principles anharmonic force constants},
    journal = {Phys. Rev. B},
    volume  = {92},
    number  = {5},
    pages   = {054301},
    year    = {2015},
    doi     = {10.1103/PhysRevB.92.054301},
}

@article{tadano2018,
    author  = {Tadano, Terumasa and Tsuneyuki, Shinji},
    title   = {First-Principles Lattice Dynamics Method for Strongly Anharmonic Crystals},
    journal = {J. Phys. Soc. Jpn.},
    volume  = {87},
    number  = {4},
    pages   = {041015},
    year    = {2018},
    doi     = {10.7566/JPSJ.87.041015},
}

@article{monacelli2021,
    author  = {Monacelli, Lorenzo and Bianco, Raffaello and Cherubini, Marco and Calandra, Matteo and Errea, Ion and Mauri, Francesco},
    title   = {The stochastic self-consistent harmonic approximation: calculating vibrational properties of materials with full quantum and anharmonic effects},
    journal = {J. Phys.: Condens. Matter},
    volume  = {33},
    number  = {36},
    pages   = {363001},
    year    = {2021},
    doi     = {10.1088/1361-648X/ac066b},
}

@article{simoncelli2019,
    author  = {Simoncelli, Michele and Marzari, Nicola and Mauri, Francesco},
    title   = {Unified theory of thermal transport in crystals and glasses},
    journal = {Nat. Phys.},
    volume  = {15},
    pages   = {809--813},
    year    = {2019},
    doi     = {10.1038/s41567-019-0520-x},
}

@article{simoncelli2022,
    author  = {Simoncelli, Michele and Marzari, Nicola and Mauri, Francesco},
    title   = {{Wigner} Formulation of Thermal Transport in Solids},
    journal = {Phys. Rev. X},
    volume  = {12},
    number  = {4},
    pages   = {041011},
    year    = {2022},
    doi     = {10.1103/PhysRevX.12.041011},
}

@article{maldonado2020,
    author  = {Maldonado, P. and Chase, T. and Reid, A. H. and Shen, X. and Li, R. K. and Carva, K. and Payer, T. and Horn von Hoegen, M. and Sokolowski-Tinten, K. and Wang, X. J. and Oppeneer, P. M. and D{\"u}rr, H. A.},
    title   = {Tracking the ultrafast nonequilibrium energy flow between electronic and lattice degrees of freedom in crystalline nickel},
    journal = {Phys. Rev. B},
    volume  = {101},
    number  = {10},
    pages   = {100302},
    year    = {2020},
    doi     = {10.1103/PhysRevB.101.100302},
}

@article{caruso2022,
    author  = {Caruso, Fabio and Novko, Dino},
    title   = {Ultrafast dynamics of electrons and phonons: from the two-temperature model to the time-dependent Boltzmann equation},
    journal = {Adv. Phys. X},
    volume  = {7},
    number  = {1},
    pages   = {2095925},
    year    = {2022},
    doi     = {10.1080/23746149.2022.2095925},
}

@article{babaei2019,
    author  = {Babaei, Hasan and Guo, Ruiqiang and Hashemi, Amirreza and Lee, Sangyeop},
    title   = {Machine-learning-based interatomic potential for phonon transport in perfect crystalline {Si} and crystalline {Si} with vacancies},
    journal = {Phys. Rev. Materials},
    volume  = {3},
    number  = {7},
    pages   = {074603},
    year    = {2019},
    doi     = {10.1103/PhysRevMaterials.3.074603},
}

@article{qian2019,
    author  = {Qian, Xin and Peng, Shenyou and Li, Xiaobo and Wei, Yujie and Yang, Ronggui},
    title   = {Thermal conductivity modeling using machine learning potentials: application to crystalline and amorphous silicon},
    journal = {Mater. Today Phys.},
    volume  = {10},
    pages   = {100140},
    year    = {2019},
    doi     = {10.1016/j.mtphys.2019.100140},
}

@article{korotaev2019,
    author  = {Korotaev, Pavel and Novoselov, Ivan and Yanilkin, Aleksey and Shapeev, Alexander},
    title   = {Accessing thermal conductivity of complex compounds by machine learning interatomic potentials},
    journal = {Phys. Rev. B},
    volume  = {100},
    number  = {14},
    pages   = {144308},
    year    = {2019},
    doi     = {10.1103/PhysRevB.100.144308},
}

@article{mortazavi2021,
    author  = {Mortazavi, Bohayra and Podryabinkin, Evgeny V. and Novikov, Ivan S. and Rabczuk, Timon and Zhuang, Xiaoying and Shapeev, Alexander V.},
    title   = {Accelerating first-principles estimation of thermal conductivity by machine-learning interatomic potentials: A {MTP/ShengBTE} solution},
    journal = {Comput. Phys. Commun.},
    volume  = {258},
    pages   = {107583},
    year    = {2021},
    doi     = {10.1016/j.cpc.2020.107583},
}

@article{fan2021,
    author  = {Fan, Zheyong and Zeng, Zezhu and Zhang, Cunzhi and Wang, Yanzhou and Song, Keke and Dong, Haikuan and Chen, Yue and Ala-Nissila, Tapio},
    title   = {Neuroevolution machine learning potentials: Combining high accuracy and low cost in atomistic simulations and application to heat transport},
    journal = {Phys. Rev. B},
    volume  = {104},
    number  = {10},
    pages   = {104309},
    year    = {2021},
    doi     = {10.1103/PhysRevB.104.104309},
}

@article{liu2021_bas_fourphonon,
    author  = {Liu, Zhichao and Yang, Xiaolong and Zhang, Bo and Li, Wu},
    title   = {High Thermal Conductivity of Wurtzite Boron Arsenide Predicted by Including Four-Phonon Scattering with Machine Learning Potential},
    journal = {ACS Appl. Mater. Interfaces},
    volume  = {13},
    number  = {45},
    pages   = {53409--53415},
    year    = {2021},
    doi     = {10.1021/acsami.1c11595},
}

@article{ouyang2022,
    author  = {Ouyang, Yulou and Yu, Cuiqian and He, Jia and Jiang, Pengfei and Ren, Weijun and Chen, Jie},
    title   = {Accurate description of high-order phonon anharmonicity and lattice thermal conductivity from molecular dynamics simulations with machine learning potential},
    journal = {Phys. Rev. B},
    volume  = {105},
    number  = {11},
    pages   = {115202},
    year    = {2022},
    doi     = {10.1103/PhysRevB.105.115202},
}

@article{togo2024_otf,
    author  = {Togo, Atsushi and Seko, Atsuto},
    title   = {On-the-fly training of polynomial machine learning potentials in computing lattice thermal conductivity},
    journal = {J. Chem. Phys.},
    volume  = {160},
    number  = {21},
    pages   = {211001},
    year    = {2024},
    doi     = {10.1063/5.0211296},
}

@article{srivastava2024,
    author  = {Srivastava, Yagyank and Jain, Ankit},
    title   = {Accelerating prediction of phonon thermal conductivity by an order of magnitude through machine learning assisted extraction of anharmonic force constants},
    journal = {Phys. Rev. B},
    volume  = {110},
    number  = {16},
    pages   = {165202},
    year    = {2024},
    doi     = {10.1103/PhysRevB.110.165202},
}

@inproceedings{batatia2022_mace,
    author    = {Batatia, Ilyes and Kov{\'a}cs, D{\'a}vid P{\'e}ter and Simm, Gregor N. C. and Ortner, Christoph and Cs{\'a}nyi, G{\'a}bor},
    title     = {{MACE}: Higher Order Equivariant Message Passing Neural Networks for Fast and Accurate Force Fields},
    booktitle = {Advances in Neural Information Processing Systems},
    volume    = {35},
    pages     = {11423--11436},
    year      = {2022},
    doi       = {10.52202/068431-0830},
}

@article{batatia2025_foundation,
  title   = {A foundation model for atomistic materials chemistry},
  author  = {Batatia, Ilyes and
             Benner, Philipp and
             Chiang, Yuan and
             Elena, Alin M. and
             Kov{\'a}cs, D{\'a}vid P. and
             Riebesell, Janosh and
             Advincula, Xavier R. and
             Asta, Mark and
             Avaylon, Matthew and
             Baldwin, William J. and
             Berger, Fabian and
             Bernstein, Noam and
             Bhowmik, Arghya and
             Bigi, Filippo and
             Blau, Samuel M. and
             C{\u a}rare, Vlad and
             Ceriotti, Michele and
             Chong, Sanggyu and
             Darby, James P. and
             De, Sandip and
             Della Pia, Flaviano and
             Deringer, Volker L. and
             Elijo{\v{s}}ius, Rokas and
             El-Machachi, Zakariya and
             Fako, Edvin and
             Falcioni, Fabio and
             Ferrari, Andrea C. and
             Gardner, John L. A. and
             Gawkowski, Miko{\l}aj J. and
             Genreith-Schriever, Annalena and
             George, Janine and
             Goodall, Rhys E. A. and
             Grandel, Jonas and
             Grey, Clare P. and
             Grigorev, Petr and
             Han, Shuang and
             Handley, Will and
             Heenen, Hendrik H. and
             Hermansson, Kersti and
             Ho, Cheuk Hin and
             Hofmann, Stephan and
             Holm, Christian and
             Jaafar, Jad and
             Jakob, Konstantin S. and
             Jung, Hyunwook and
             Kapil, Venkat and
             Kaplan, Aaron D. and
             Karimitari, Nima and
             Kermode, James R. and
             Kourtis, Panagiotis and
             Kroupa, Namu and
             Kullgren, Jolla and
             Kuner, Matthew C. and
             Kuryla, Domantas and
             Liepuoniute, Guoda and
             Lin, Chen and
             Margraf, Johannes T. and
             Magd{\u{a}}u, Ioan-Bogdan and
             Michaelides, Angelos and
             Moore, J. Harry and
             Naik, Aakash A. and
             Niblett, Samuel P. and
             Norwood, Sam Walton and
             O'Neill, Niamh and
             Ortner, Christoph and
             Persson, Kristin A. and
             Reuter, Karsten and
             Rosen, Andrew S. and
             Rosset, Louise A. M. and
             Schaaf, Lars L. and
             Schran, Christoph and
             Shi, Benjamin X. and
             Sivonxay, Eric and
             Stenczel, Tam{\'a}s K. and
             Sutton, Christopher and
             Svahn, Viktor and
             Swinburne, Thomas D. and
             Tilly, Jules and
             van der Oord, Cas and
             Vargas, Santiago and
             Varga-Umbrich, Eszter and
             Vegge, Tejs and
             Vondr{\'a}k, Martin and
             Wang, Yangshuai and
             Witt, William C. and
             Wolf, Thomas and
             Zills, Fabian and
             Cs{\'a}nyi, G{\'a}bor},
  journal = {The Journal of Chemical Physics},
  volume  = {163},
  number  = {18},
  pages   = {184110},
  year    = {2025},
  doi     = {10.1063/5.0297006}
}

@article{loew2025,
    author  = {Loew, Antoine and Sun, Dewen and Wang, Hai-Chen and Botti, Silvana and Marques, Miguel A. L.},
    title   = {Universal machine learning interatomic potentials are ready for phonons},
    journal = {npj Comput. Mater.},
    volume  = {11},
    pages   = {178},
    year    = {2025},
    doi     = {10.1038/s41524-025-01650-1},
}

@misc{kaplan2025_matpes,
    author        = {Kaplan, Aaron D. and Liu, Runze and Qi, Ji and Ko, Tsz Wai and Deng, Bowen and Riebesell, Janosh and Ceder, Gerbrand and Persson, Kristin A. and Ong, Shyue Ping},
    title         = {A Foundational Potential Energy Surface Dataset for Materials},
    year          = {2025},
    eprint        = {2503.04070},
    archivePrefix = {arXiv},
    primaryClass  = {cond-mat.mtrl-sci},
    doi           = {10.48550/arXiv.2503.04070},
}

@inproceedings{koker2026_pft,
  title     = {{PFT}: Phonon Fine-tuning for Machine Learned Interatomic Potentials},
  author    = {Koker, Teddy and Gangan, Abhijeet Sadashiv and Kotak, Mit and Marian, Jaime and Smidt, Tess},
  booktitle = {Proceedings of the 43rd International Conference on Machine Learning},
  series    = {Proceedings of Machine Learning Research},
  volume    = {306},
  pages     = {59932--59948},
  year      = {2026},
  publisher = {PMLR},
  url       = {https://proceedings.mlr.press/v306/koker26a.html}
}

@article{kaldo2026,
    author  = {Barbalinardo, Giuseppe and Chen, Zekun and Folkner, Dylan A. and Li, Bohan and Lundgren, Nicholas W. and Troup, Nathaniel and Fiorentino, Alfredo and Donadio, Davide},
    title   = {{$\kappa$ALDo 2.0}: Scalable thermal transport from first principles and machine learning potentials},
    journal = {Comput. Phys. Commun.},
    volume  = {327},
    pages   = {110282},
    year    = {2026},
    doi     = {10.1016/j.cpc.2026.110282},
}

@article{qin2026_mlp_review,
    author  = {Qin, Yangjun and Zong, Zhicheng and Li, Tianhao and Fang, Haisheng and Zhu, Guimei and Jiang, Jinwu and Yang, Nuo},
    title   = {Studying Heat Conduction Properties by Using Machine Learning Potentials: a Brief Review},
    journal = {Chin. Phys. Lett.},
    volume  = {43},
    number  = {8},
    pages   = {080802},
    year    = {2026},
    doi     = {10.1088/0256-307X/43/8/080802},
}

@article{liu2025_nep_specific_heat,
    author  = {Liu, Shixian and Zhang, Ge and Yin, Fei and Barinov, A. A. and Khvesyuk, V. I. and Yang, Nuo},
    title   = {Temperature dependence of specific heat capacity of nanostructures via neuroevolution machine-learned potential},
    journal = {J. Appl. Phys.},
    volume  = {138},
    pages   = {104301},
    year    = {2025},
    doi     = {10.1063/5.0284002},
}

@article{yin2026_nepkappa,
    author  = {Yin, Fei and Liu, Shixian and Dong, Yiming and Barinov, A. A. and Xu, Ke and Khvesyuk, V. I.},
    title   = {Accelerated phonon transport calculations for nanostructures: Combining neuroevolution potentials and compressed sensing},
    journal = {J. Appl. Phys.},
    volume  = {139},
    pages   = {135103},
    year    = {2026},
    doi     = {10.1063/5.0324012},
}

@article{liang2026_nep89,
  title   = {{NEP89}: universal neuroevolution potential for inorganic and organic materials across 89 elements},
  author  = {Liang, Ting and Xu, Ke and Lindgren, Eric and
             Chen, Zherui and Zhao, Rui and Liu, Jiahui and
             Berger, Esm{\'e}e and Tang, Benrui and
             Zhang, Bohan and Wang, Yanzhou and
             Song, Keke and Ying, Penghua and Xu, Nan and
             Dong, Haikuan and Chen, Shunda and
             Erhart, Paul and Fan, Zheyong and
             Ala-Nissila, Tapio and Xu, Jianbin},
  journal = {Nature Computational Science},
  volume  = {6},
  number  = {7},
  pages   = {789--801},
  year    = {2026},
  doi     = {10.1038/s43588-026-01009-6}
}

@article{belli2025_sscha_mlip,
  title   = {Efficient modelling of anharmonicity and quantum effects in {PdCuH$_2$} with machine learning potentials},
  author  = {Belli, Francesco and Zurek, Eva},
  journal = {npj Computational Materials},
  volume  = {11},
  pages   = {87},
  year    = {2025},
  doi     = {10.1038/s41524-025-01553-1}
}

@article{lee2025_highthroughput,
  title   = {Accelerating high-throughput phonon calculations via machine learning universal potentials},
  author  = {Lee, Huiju and Hegde, Vinay I. and Wolverton, Chris and Xia, Yi},
  journal = {Materials Today Physics},
  volume  = {53},
  pages   = {101688},
  year    = {2025},
  doi     = {10.1016/j.mtphys.2025.101688}
}

@book{born1954,
  title     = {Dynamical Theory of Crystal Lattices},
  author    = {Born, Max and Huang, Kun},
  publisher = {Oxford University Press},
  address   = {Oxford},
  year      = {1954}
}

@book{srivastava2023,
  title     = {The Physics of Phonons},
  author    = {Srivastava, Gyaneshwar P.},
  edition   = {2},
  publisher = {CRC Press},
  year      = {2023},
  isbn      = {9780367693169},
  doi       = {10.1201/9781003141273}
}

@article{bianco2017,
  title   = {Second-order structural phase transitions, free energy curvature, and temperature-dependent anharmonic phonons in the self-consistent harmonic approximation: Theory and stochastic implementation},
  author  = {Bianco, Raffaello and Errea, Ion and Paulatto, Lorenzo and Calandra, Matteo and Mauri, Francesco},
  journal = {Physical Review B},
  volume  = {96},
  pages   = {014111},
  year    = {2017},
  doi     = {10.1103/PhysRevB.96.014111}
}

@article{feng2014prediction,
  title   = {Prediction of Spectral Phonon Mean Free Path and Thermal Conductivity with Applications to Thermoelectrics and Thermal Management: A Review},
  author  = {Feng, Tianli and Ruan, Xiulin},
  journal = {Journal of Nanomaterials},
  volume  = {2014},
  pages   = {206370},
  year    = {2014},
  doi     = {10.1155/2014/206370}
}

@article{klemens1959,
  title   = {Thermal resistance due to isotopes and other point defects},
  author  = {Klemens, P. G.},
  journal = {Journal of Physics and Chemistry of Solids},
  volume  = {8},
  pages   = {345--347},
  year    = {1959},
  doi     = {10.1016/0022-3697(59)90359-2}
}

@article{tamura1983,
  title   = {Isotope scattering of dispersive phonons in {Ge}},
  author  = {Tamura, Shinichiro},
  journal = {Physical Review B},
  volume  = {27},
  number  = {2},
  pages   = {858--866},
  year    = {1983},
  doi     = {10.1103/PhysRevB.27.858}
}

@article{ono2017,
  title   = {Nonequilibrium phonon dynamics beyond the quasiequilibrium approach},
  author  = {Ono, Shota},
  journal = {Physical Review B},
  volume  = {96},
  pages   = {024301},
  year    = {2017},
  doi     = {10.1103/PhysRevB.96.024301}
}

@article{wu2024_forceerror,
  title   = {Correcting force error-induced underestimation of lattice thermal conductivity in machine learning molecular dynamics},
  author  = {Wu, Xiguang and Zhou, Wenjiang and Dong, Haikuan and Ying, Penghua and Wang, Yanzhou and Song, Bai and Fan, Zheyong and Xiong, Shiyun},
  journal = {The Journal of Chemical Physics},
  volume  = {161},
  number  = {1},
  pages   = {014103},
  year    = {2024},
  doi     = {10.1063/5.0213811}
}

@article{olego1982_sic_decay,
  title   = {Pressure dependence of Raman phonons of {Ge} and {3C-SiC}},
  author  = {Olego, Diego and Cardona, Manuel},
  journal = {Physical Review B},
  volume  = {25},
  number  = {2},
  pages   = {1151--1160},
  year    = {1982},
  doi     = {10.1103/PhysRevB.25.1151}
}

@article{matulionis2004_sic_hotphonon,
  title   = {Hot-phonon temperature and lifetime in biased {4H-SiC}},
  author  = {Matulionis, A. and Liberis, J. and Matulionien{\.e}, I. and Cha, H.-Y. and Eastman, L. F. and Spencer, M. G.},
  journal = {Journal of Applied Physics},
  volume  = {96},
  number  = {11},
  pages   = {6439--6444},
  year    = {2004},
  doi     = {10.1063/1.1812598}
}

@article{yoshida2013_sic_excitation,
  title   = {Experimental demonstration of mode-selective phonon excitation of {6H-SiC} by a mid-infrared laser with anti-Stokes Raman scattering spectroscopy},
  author  = {Yoshida, Kyohei and Sonobe, Taro and Zen, Heishun and Hachiya, Kan and Okumura, Kensuke and Mishima, Kenta and Inukai, Motoharu and Negm, Hani and Torgasin, Konstantin and Omer, Mohamed and Kii, Toshiteru and Masuda, Kai and Ohgaki, Hideaki},
  journal = {Applied Physics Letters},
  volume  = {103},
  number  = {18},
  pages   = {182103},
  year    = {2013},
  doi     = {10.1063/1.4827253}
}

@article{xu2025gpumd,
  title   = {{GPUMD} 4.0: A high-performance molecular dynamics package for versatile materials simulations with machine-learned potentials},
  author  = {Xu, Ke and
             Bu, Hekai and
             Pan, Shuning and
             Lindgren, Eric and
             Wu, Yongchao and
             Wang, Yong and
             Liu, Jiahui and
             Song, Keke and
             Xu, Bin and
             Li, Yifan and
             Hainer, Tobias and
             Svensson, Lucas and
             Wiktor, Julia and
             Zhao, Rui and
             Huang, Hongfu and
             Qian, Cheng and
             Zhang, Shuo and
             Zeng, Zezhu and
             Zhang, Bohan and
             Tang, Benrui and
             Xiao, Yang and
             Yan, Zihan and
             Shi, Jiuyang and
             Liang, Zhixin and
             Wang, Junjie and
             Liang, Ting and
             Cao, Shuo and
             Wang, Yanzhou and
             Ying, Penghua and
             Xu, Nan and
             Chen, Chengbing and
             Zhang, Yuwen and
             Chen, Zherui and
             Wu, Xin and
             Jiang, Wenwu and
             Berger, Esm{\'e}e and
             Li, Yanlong and
             Chen, Shunda and
             Gabourie, Alexander J. and
             Dong, Haikuan and
             Xiong, Shiyun and
             Wei, Ning and
             Chen, Yue and
             Xu, Jianbin and
             Ding, Feng and
             Sun, Zhimei and
             Ala-Nissila, Tapio and
             Harju, Ari and
             Zheng, Jincheng and
             Guan, Pengfei and
             Erhart, Paul and
             Sun, Jian and
             Ouyang, Wengen and
             Su, Yanjing and
             Fan, Zheyong},
  journal = {Materials Genome Engineering Advances},
  volume  = {3},
  number  = {3},
  pages   = {e70028},
  year    = {2025},
  doi     = {10.1002/mgea.70028}
}

@article{zhao2026_in2se3,
  title   = {Polarization-driven enhancement of the thermal switch ratio arising from strong four-phonon scattering in bilayer {$\alpha$-In$_2$Se$_3$}},
  author  = {Zhao, Zhuo and Zhang, Jian and Zhou, Wu-Xing and Cheng, Yuan and Zhang, Gang},
  journal = {Nanoscale},
  volume  = {18},
  number  = {33},
  pages   = {17703--17710},
  year    = {2026},
  doi     = {10.1039/D6NR01935H}
}

@article{zhang2026_mnse,
  title   = {Ultrafast thermal conductivity switching in monolayer {MnSe} triggered by laser-induced intersite spin transfer},
  author  = {Zhang, Dingbo and Li, Shuo and Chen, Wenwen and Ni, Yuxiang and Zhang, Gang},
  journal = {Physical Review B},
  volume  = {113},
  pages   = {195437},
  year    = {2026},
  doi     = {10.1103/rr2t-mqq7}
}

@article{spohn2006,
  title   = {The phonon Boltzmann equation, properties and link to weakly anharmonic lattice dynamics},
  author  = {Spohn, Herbert},
  journal = {Journal of Statistical Physics},
  volume  = {124},
  pages   = {1041--1104},
  year    = {2006},
  doi     = {10.1007/s10955-005-8088-5}
}

@article{yao2025_rtbte,
  title   = {Advancing simulations of coupled electron and phonon nonequilibrium dynamics using adaptive and multirate time integration},
  author  = {Yao, Jia and Maliyov, Ivan and Gardner, David J. and Woodward, Carol S. and Bernardi, Marco},
  journal = {npj Computational Materials},
  volume  = {11},
  pages   = {256},
  year    = {2025},
  doi     = {10.1038/s41524-025-01738-8}
}

@article{yin_cove_2027,
	title = {Cove-edge-induced vibrational confinement and weakened collective phonon transport in graphene nanoribbons},
	volume = {233},
	doi = {10.1016/j.ijthermalsci.2026.111381},
	journal = {International Journal of Thermal Sciences},
	author = {Yin, Fei and Wang, Gang and Barinov, A.A. and Liu, Shixian and Xu, Ke},
	year = {2027},
	pages = {111381},
}

@article{han2023graphene,
  title = {Thermal conductivity of monolayer graphene: Convergent and lower than diamond},
  author = {Han, Zherui and Ruan, Xiulin},
  journal = {Physical Review B},
  volume = {108},
  number = {12},
  pages = {L121412},
  year = {2023},
  doi = {10.1103/PhysRevB.108.L121412}
}

@article{li2024convergent,
  title = {Convergent thermal conductivity in strained monolayer graphene},
  author = {Li, Guotai and Tang, Jialin and Zheng, Jiongzhi
            and Wang, Qi and Cui, Zheng and Xu, Ke
            and Xu, Jianbin and Liu, Te-Huan and Zhu, Guimei
            and Guo, Ruiqiang and Li, Baowen},
  journal = {Physical Review B},
  volume = {109},
  number = {3},
  pages = {035420},
  year = {2024},
  doi = {10.1103/PhysRevB.109.035420}
}

@article{jain2024fourphonon,
  title = {Impact of four-phonon scattering on thermal transport in carbon nanotubes},
  author = {Jain, Ankit},
  journal = {Physical Review B},
  volume = {109},
  number = {15},
  pages = {155413},
  year = {2024},
  doi = {10.1103/PhysRevB.109.155413}
}

@article{ravichandran2019nonmonotonic,
  title = {Non-monotonic pressure dependence of the thermal conductivity of boron arsenide},
  author = {Ravichandran, Navaneetha K. and Broido, David},
  journal = {Nature Communications},
  volume = {10},
  pages = {827},
  year = {2019},
  doi = {10.1038/s41467-019-08713-0}
}

@article{dai2025machine,
  title = {Machine learning for thermal transport and phonon high-order anharmonicity in high thermal conductivity materials: A case study in boron arsenide},
  author = {Dai, Lingyun and Li, Man and Hu, Yongjie},
  journal = {Physical Review Materials},
  volume = {9},
  number = {4},
  pages = {045403},
  year = {2025},
  doi = {10.1103/PhysRevMaterials.9.045403}
}

@article{li_higher-order-phonon_2026,
	title = {Higher-order-phonon scattering governs targeted control of heat conduction in bulk boron arsenide},
	volume = {68},
	doi = {10.1016/j.mtphys.2026.102236},
	urldate = {2026-09-26},
	journal = {Materials Today Physics},
	author = {Li, Tianhao and Qin, Yangjun and Pan, Dongkai and Liu, Shixian and Meng, Han and Yang, Nuo},
	year = {2026},
	pages = {102236},
}

@article{shapeev2016_mtp,
  title   = {Moment Tensor Potentials: A Class of Systematically Improvable Interatomic Potentials},
  author  = {Shapeev, Alexander V.},
  journal = {Multiscale Modeling \& Simulation},
  volume  = {14},
  number  = {3},
  pages   = {1153--1173},
  year    = {2016},
  doi     = {10.1137/15M1054183}
}

@article{bartok2010,
  title   = {Gaussian Approximation Potentials: The Accuracy of Quantum Mechanics, without the Electrons},
  author  = {Bart{\'o}k, Albert P. and Payne, Mike C. and Kondor, Risi and Cs{\'a}nyi, G{\'a}bor},
  journal = {Physical Review Letters},
  volume  = {104},
  number  = {13},
  pages   = {136403},
  year    = {2010},
  doi     = {10.1103/PhysRevLett.104.136403}
}

@article{ladygin2020,
  title   = {Lattice dynamics simulation using machine learning interatomic potentials},
  author  = {Ladygin, V. V. and Korotaev, P. Yu. and Yanilkin, A. V. and Shapeev, A. V.},
  journal = {Computational Materials Science},
  volume  = {172},
  pages   = {109333},
  year    = {2020},
  doi     = {10.1016/j.commatsci.2019.109333}
}

@article{choi2022_nn,
  title   = {Accelerated computation of lattice thermal conductivity using neural network interatomic potentials},
  author  = {Choi, Jeong Min and Lee, Kyeongpung and Kim, Sangtae and Moon, Minseok and Jeong, Wonseok and Han, Seungwu},
  journal = {Computational Materials Science},
  volume  = {211},
  pages   = {111472},
  year    = {2022},
  doi     = {10.1016/j.commatsci.2022.111472}
}

@article{wyant2021_interface,
  title   = {Machine learned interatomic potentials for modeling interfacial heat transport in {Ge/GaAs}},
  author  = {Wyant, Spencer and Rohskopf, Andrew and Henry, Asegun},
  journal = {Computational Materials Science},
  volume  = {200},
  pages   = {110836},
  year    = {2021},
  doi     = {10.1016/j.commatsci.2021.110836}
}

@article{tang2023_hbn,
  title   = {Effect of four-phonon scattering on anisotropic thermal transport in bulk hexagonal boron nitride by machine learning interatomic potential},
  author  = {Tang, Jialin and Li, Guotai and Wang, Qi and Zheng, Jiongzhi and Cheng, Lin and Guo, Ruiqiang},
  journal = {International Journal of Heat and Mass Transfer},
  volume  = {207},
  pages   = {124011},
  year    = {2023},
  doi     = {10.1016/j.ijheatmasstransfer.2023.124011}
}

\end{document}